\documentclass{article}
\usepackage[english]{babel}
\usepackage{amsmath,amssymb,amsfonts,graphicx,xcolor}
\usepackage{fullpage}
\usepackage{url}

\newcommand{\assign}{:=}
\newcommand{\asterisk}{\mathord{*}}
\newcommand{\mathd}{\mathrm{d}}
\newcommand{\mathpi}{\pi}
\newcommand{\nocomma}{}
\newcommand{\of}{:}
\newcommand{\tmem}[1]{{\em #1\/}}
\newcommand{\tmmathbf}[1]{\ensuremath{\boldsymbol{#1}}}
\newcommand{\tmop}[1]{\ensuremath{\operatorname{#1}}}
\newcommand{\tmstrong}[1]{\textbf{#1}}
\newcommand{\tmverbatim}[1]{\text{{\ttfamily{#1}}}}

\newcommand{\tmaffiliation}[1]{\\ #1}
\numberwithin{equation}{section}
\numberwithin{figure}{section}
\usepackage{hyperref}

\newcommand{\tmmult}{\,} % space for multiplication

\title{Simple deformation measures\\ for Discrete elastic rods and ribbons}

\author{
  Kevin Korner
  \tmaffiliation{Division of Engineering and Applied Science\\
  California Institute of Technology\\
  Pasadena, CA 91125, USA}
  \and
  Basile Audoly
  \tmaffiliation{Laboratoire de M{\'e}canique des Solides\\
  CNRS, Institut Polytechnique de Paris\\
  91120 Palaiseau, France}
  \and
  Kaushik Bhattacharya
  \tmaffiliation{Division of Engineering and Applied Science\\
  California Institute of Technology\\
  Pasadena, CA 91125, USA}
}

\date{November 4, 2021}

\begin{document}

\maketitle

\begin{abstract}
  The Discrete elastic rod method (Bergou {\tmem{et al.}}, 2008) is a
  numerical method for simulating slender elastic bodies.  It works
  by representing the center-line as a polygonal chain,
  attaching two perpendicular directors to each segment,
  and 
  defining discrete stretching, bending and twisting deformation
  measures and a discrete strain energy.  Here, we investigate an
  alternative formulation of this model based on a simpler definition
  of the discrete deformation measures.  Both formulations are equally consistent
  with the continuous rod model.  Simple formulas for the first and
  second gradients of the discrete deformation measures are derived,
  making it easy to calculate the Hessian of the discrete strain
  energy.  A few numerical illustrations are given.  The approach is
  also extended to inextensible ribbons described by
  the Wunderlich model, and both the developability constraint and the
  dependence of the energy of the strain gradients are handled
  naturally.
\end{abstract}

\section{Introduction}

The geometric non-linearity of thin elastic rods gives rise to a rich
range of phenomena even when the strains are small, see e.g.\
{\cite{Baek-Reis-Rigidity-of-hemispherical-elastic-2019,Panetta-Konakovic-Lukovic-EtAl-X-Shells:-A-New-Class-of-Deployable-2019}}
for recent examples.  So, the non-linear theory of rods has
traditionally combined geometrically non-linearity with linear
constitutive laws~\cite{Antman,Audoly-Pomeau-Elasticity-and-geometry:-from-2010}.
However, recent interest has expanded beyond the linearly elastic
regime, including viscous threads
{\cite{Brun-Ribe-EtAl-A-numerical-investigation-of-the-fluid-2012,Ribe-Habibi-EtAl-Liquid-rope-coiling-2012}},
plastic and visco-plastic
bars~{\cite{Coleman-Newman-On-the-rheology-of-cold-drawing.-1988,Audoly-Hutchinson-Analysis-of-necking-based-2016,Audoly-Hutchinson-One-dimensional-modeling-of-necking-2019}},
visco-elastic rods
{\cite{Lestringant-Audoly-EtAl-A-discrete-geometrically-exact-2020}},
capillary elastic beams made of very soft
materials~{\cite{Lestringant-Audoly-A-one-dimensional-model-for-elasto-capillary-2020}}.
Thin elastic ribbons may also be viewed in this class with a
non-linear constitutive law that captures the complex deformation of
the
cross-sections~{\cite{Sadowsky-Die-Differentialgleichungen-des-Mobiusschen-Bandes-1929,Wunderlich-Uber-ein-abwickelbares-Mobiusband-1962,Starostin-Heijden-The-shape-of-a-Mobius-strip-2007,Starostin-Heijden-Equilibrium-Shapes-with-2015,Dias-Audoly-Wunderlich-meet-Kirchhoff:-2015,Audoly-Neukirch-A-one-dimensional-model-for-elastic-2021}}.

The study of instabilities, especially in the presence of complex constitutive relations, requires an accurate but efficient numerical method.  
Here, we build on the work of Bergou {\it et al.} \cite{Bergou-Wardetzky-EtAl-Discrete-Elastic-Rods-2008}
to propose a numerical method applicable to slender elastic structures
in general. 
To keep the presentation focused, we limit our presentation to
{\tmem{elastic}} rods: both linearly elastic and non-linear elastic
constitutive laws are covered. Our main contribution consists in providing a
discrete geometric description of slender rods. This kinematic building block is
independent of the elastic constitutive law in our formulation, making the
extension to inelastic constitutive laws relatively straightforward, as
discussed in Section~\ref{sec:ribbons}.

We follow the classical kinematic approach, and use the  arc-length $s$ in the undeformed 
configuration as a Lagrangian coordinate.  We denote the center-line of the rod in
the current configuration as $\tmmathbf{x} (s)$ (boldface symbols denote vectors).  
We introduce an orthonormal
set of vectors $(\tmmathbf{d}_I (s))_{1 \leqslant I \leqslant 3}$, called the
{\tmem{directors}}, to describe the orientation of the cross-section.  We 
impose the {\tmem{adaptation condition}} that the
director $\tmmathbf{d}_3$ matches the unit tangent $\tmmathbf{t}$ to the
center-line:
\begin{equation}
  \tmmathbf{d}_3 (s) =\tmmathbf{t} (s), \text{ where $\tmmathbf{t} (s) =
  \frac{\tmmathbf{x}' (s)}{|\tmmathbf{x}' (s) |}$.}
  \label{eq:adaptation-continuous}
\end{equation}
Here $\tmmathbf{x}' (s) = \partial \tmmathbf{x}/ \partial s$ denotes
the derivative of $\tmmathbf{x}$ with respect to the arc-length $s$.  Note that the
adaptation condition does {\tmem{not}} impose any restriction on the
actual deformation of the rod at the microscopic scale; specifically, it
does not require the deformed cross-section to be spanned by
$\tmmathbf{d}_1$ and $\tmmathbf{d}_2$.  Instead, it expresses the fact
that the only role of the directors is to track the twisting
motion of the cross-sections about the tangent.
Equation~(\ref{eq:adaptation-continuous}) does not impose
inextensibility either.

The rotation gradient
$\tmmathbf{\kappa} (s)$, also known as the Darboux vector, is defined by
\begin{equation}
  \tmmathbf{d}_I' (s) =\tmmathbf{\kappa} (s) \times \tmmathbf{d}_I (s), \quad \quad I = 1, 2, 3.
  \label{eq:rotation-gradient-gradient}
\end{equation}
It exists and is unique since the directors are
orthonormal. The deformation measures are
\begin{equation}
  \kappa_{(I)} (s) =\tmmathbf{\kappa} (s) \cdot \tmmathbf{d}_I (s)
  \label{eq:k-sub-I-continuous}
\end{equation}
A fourth deformation measure is introduced to
characterize how the center-line stretches, such as $\varepsilon (s) =
\frac{1}{2} \tmmult \left( {\tmmathbf{x}'}^2 (s) - 1 \right)$ (Green-Lagrange
strain).

This kinematic description is common to all variants of the rod model.
It is complemented by constitutive equations specifying either the
stored energy density (in the case of a hyperelastic theory) or the
reaction forces and moments as functions of the four deformation
measures or their histories.  The formulation is completed by imposing
either equilibrium or balance of momenta.  The resulting equations for
linear elastic constitutive relations are known as the Kirchhoff
equations for rods, and they can be derived variationally, see
~{\cite{Steigmann-Faulkner-Variational-theory-for-spatial-1993,Audoly-Pomeau-Elasticity-and-geometry:-from-2010}};
we will not discuss them further.
we will not discuss them further.

Various strategies have been proposed to simulate the equations for thin rods numerically. 
In approaches based on the finite-element methods, it is
challenging to represent the kinematic constraint of
adaptation~(\ref{eq:adaptation-continuous}) between the unknown center-line
$\tmmathbf{x} (s)$ and the unknown rotation representing the orthonormal
directors $\tmmathbf{d}_I (s)$. Another approach is based on super-helices or
super-clothoids: in these high-order approaches, the bending and twisting
strain measures $\kappa_{(I)} (s)$ are discretized into constant or piecewise
linear functions. The result is a highly accurate method which has been
successfully applied to several challenging
problems~{\cite{Bertails-Audoly-EtAl-Super-Helices-for-Predicting-the-Dynamics-2006,Casati-Bertails-Descoubes-Super-Space-Clothoids-2013,Charrondiere-Bertails-Descoubes-EtAl-Numerical-modeling-of-inextensible-2020}}.
The price to pay is that the reconstruction of the center-line in
terms of the degrees of freedom is non-trivial and non-local. Additionally, some
common boundary conditions, such as clamped-clamped conditions, must be
treated using non-linear constraints.

\begin{figure}
  \centerline{\includegraphics[width=.99\textwidth]{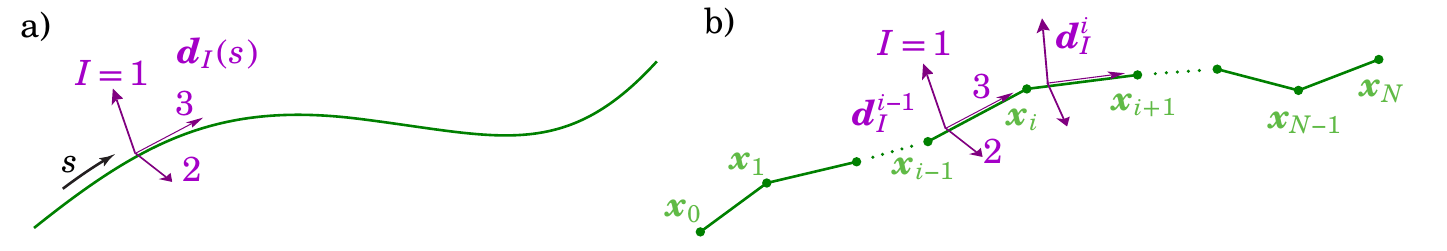}}
  \caption{(a)~A continuous elastic rod and (b)~a discrete elastic rod. The
  adaptation condition from equations~(\ref{eq:adaptation-continuous})
  and~(\ref{eq:adaptation-constraint-current}) is satisfied in both
  cases.\label{fig:continuous-vs-discrete}}
\end{figure}

A new approach called the Discrete elastic rods method was introduced by 
Bergou {\it et al.} \cite{Bergou-Wardetzky-EtAl-Discrete-Elastic-Rods-2008};
see {\cite{Jawed-Novelia-EtAl-A-primer-on-the-kinematics-of-discrete-2017}}
for a recent primer.
The Discrete elastic rod method is a low-order method, which starts
out by discretizing the center-line into a polygonal chain with nodes
$(\tmmathbf{x}_0, \ldots, \tmmathbf{x}_N)$.  The tangents and material
frames $\tmmathbf{d}_I^i$ are defined on the segments, see
Figure~\ref{fig:continuous-vs-discrete}.  The adaptation
condition~(\ref{eq:adaptation-continuous}) is used to parameterize the
material frames $(\tmmathbf{d}_I^i)_{1 \leqslant i \leqslant 3}$ in
terms of the positions $(\tmmathbf{x}_{i - 1}, \tmmathbf{x}_i)$ of the
adjacent nodes and of a single twisting angle $\varphi^i$, as described
in Section~\ref{sec:discrete-kinematics}\ref{ssec:centerline-twist}.
A discrete rotation gradient is obtained by comparing the orthonormal
directors from adjacent segments: this yields a differential rotation
 at a  {\tmem{vertex}} between the segments.  This must now be projected
onto a material frame to yield the bending and twisting strain
measures according to equation~(\ref{eq:k-sub-I-continuous}).  The material
frame, however, lives on {\tmem{segments}}.  The original Discrete
elastic rod formulation worked around this difficulty by introducing
an additional director frame living on the nodes, obtained by
averaging the director frames from the adjacent
segments~{\cite{Bergou-Wardetzky-EtAl-Discrete-Elastic-Rods-2008,Jawed-Novelia-EtAl-A-primer-on-the-kinematics-of-discrete-2017}}.
In the present work, a different definition of the discrete bending
and twisting strain measures is used, see
Equations~(\ref{eq:rotation-gradient}) and~(\ref{eq:kappi}).  This small
change simplifies the formulation of model considerably.  We note
that a similar measure was introduced independently in a recent work
on shearable rod
models~{\cite{Gazzola-Dudte-EtAl-Forward-and-inverse-problems-2018}}.

Overall, the proposed formulation offers the following advantages:
\begin{itemize}
  \item As in the original Discrete rod model, the proposed formulation eliminates two out of
  the three degrees of freedom associated with the directors at
  each node using of the adaptation condition~(\ref{eq:adaptation-continuous}); this leads to a
  constraint-free formulation that uses degrees of freedom sparingly.

  \item The formulation of the model is concise: in particular the
  gradient and Hessian of the discrete elastic energy are given
  by the simple, closed form formulas listed in Section~\ref{sec:variations}.
  
  \item The proposed deformation measures have a clear geometric interpretation: in
  the context of inextensible ribbons, for example, a discrete developability
  condition can easily be formulated in terms of the new set of discrete
  strains, see Section~\ref{sec:discrete-kinematics}\ref{ssec:ribbons}.
  
  \item The kinematic description can easily be combined with various constitutive models to produce
discrete models for elastic rods, inextensible ribbons, viscous or
visco-elastic rods, etc., as discussed in Section~\ref{sec:ribbons}.
  
\end{itemize}

\section{Discrete bending and twisting deformation measures}
\label{sec:discrete-kinematics}

\subsection{A compendium on quaternions}

Rod models make use of rotations in the three-dimensional space.
These rotations are conveniently represented using quaternions.  Here,
we provide a brief summary of quaternions and their main properties.
A complete and elementary introduction to quaternions can be found 
in \cite{Morais2014}.

A quaternion $q \in \mathbb{Q}$ can be seen as a pair made up of a scalar $s
\in \mathbb{R}$ and a vector $\tmmathbf{v} \in \mathbb{R}^3$, $q = (s,
\tmmathbf{v})$. Identifying the scalar $s$ and the vector $\tmmathbf{v}$ with
the quaternions $(s, \tmmathbf{0})$ and $(0, \tmmathbf{v})$ respectively, one
has the quaternion decomposition
\[ q = s +\tmmathbf{v}. \]
The product of two quaternions $q_1 = (s_1, \tmmathbf{v}_1)$ and $q_2 = (s_2,
\tmmathbf{v}_2)$ is defined as
\begin{equation}
  q_1 \tmmult q_2 = \left( s_1 \tmmult s_2 -\tmmathbf{v}_1 \cdot
  \tmmathbf{v}_2 \right) + (s_1 \tmmathbf{v}_2 + s_2 \tmmathbf{v}_1
  +\tmmathbf{v}_1 \times \tmmathbf{v}_2) . \label{eq:quaternion-mult}
\end{equation}
The product is non-commutative.

A unit quaternion $r = s +\tmmathbf{v}$ is a quaternion such that $s^2 + |
\tmmathbf{v} |^2 = 1$. Unit quaternions represent rotations in the
three-dimensional Euclidean space, in the following sense. Define $\bar{r} = s
-\tmmathbf{v}$ as the quaternion conjugate to $r$. Define the action of the
unit quaternion $r$ on an arbitrary vector $\tmmathbf{w}$ as
\[ r \ast \tmmathbf{w}= r \tmmult \tmmathbf{w} \tmmult \bar{r}, \]
where the left-hand side defines a linear map on the set of vectors
$\tmmathbf{w}$, and the right-hand side is a double product of quaternions. It
can be shown that ({\tmem{i)}}~the quaternion $r \ast \tmmathbf{w}$ is a pure
vector, ({\tmem{ii}})~the mapping $\tmmathbf{w} \to r \ast \tmmathbf{w}$ is a
rotation in Euclidean space, ({\tmem{iii}})~the quaternion $r$ can be written
as $r = \pm r_{\tmmathbf{n}} (\theta)$ where
\begin{equation}
  r_{\tmmathbf{n}} (\theta) = \cos \frac{\theta}{2} +\tmmathbf{n} \tmmult \sin
  \frac{\theta}{2} = \exp \frac{\tmmathbf{n} \tmmult \theta}{2},
  \label{eq:unit-quaternion-normal-form}
\end{equation}
$\theta$ is the angle of the rotation, and $\tmmathbf{n}$ is a unit vector
subtending the axis of the rotation. Note that both unit quaternions $+
r_{\tmmathbf{n}} (\theta)$ and $- r_{\tmmathbf{n}} (\theta)$ represent the
same rotation.

Given two unit quaternions $r_1$ and $r_2$, consider the product $r_2 \tmmult
r_1$: for any vector $\tmmathbf{w}$, the equality $\left( r_2 \tmmult r_1
\right) \ast \tmmathbf{w}= r_2 \tmmult r_1 \tmmult \tmmathbf{w} \tmmult
\overline{r_2 \tmmult r_1} = r_2 \tmmult r_1 \tmmult \tmmathbf{w} \tmmult
\overline{r_1} \tmmult \overline{r_2} = r_2 \ast (r_1 \ast \tmmathbf{w})$
shows that the unit quaternion $r_2 \tmmult r_1$ represents the
{\tmem{composition}} of the rotations associated with $r_1$ applied first,
and $r_2$ applied last. The multiplication of unit quaternions is
therefore equivalent to the composition of rotations. In view of this, we will
{\tmem{identify}} rotations with unit quaternions. The inverse of the rotation
$r$ will accordingly be identified with the conjugate $\overline{r}$. 

\subsection{Parallel transport}\label{ssec:parallel_transport}

Parallel transport plays a key role in the Discrete elastic rods model, by
allowing one to define twistless configurations of the material frame in an
intrinsic way. For two unit vectors $\tmmathbf{a}$ and $\tmmathbf{b}$ such
that $\tmmathbf{b} \neq -\tmmathbf{a}$, the parallel transport from
$\tmmathbf{a}$ to $\tmmathbf{b}$ is the rotation mapping $\tmmathbf{a}$ to
$\tmmathbf{b}$, whose axis is along the binormal $\tmmathbf{a} \times
\tmmathbf{b}$.  Parallel transport can be interpreted geometrically as the
rotation mapping $\tmmathbf{a}$ to $\tmmathbf{b}$ and tracing out the shortest
path on the unit
sphere~{\cite{Bergou-Wardetzky-EtAl-Discrete-Elastic-Rods-2008}}.

An explicit expression of the parallel transport from $\tmmathbf{a}$ to
$\tmmathbf{b}$ in terms of unit quaternions
is~{\cite{Linn-Discrete-Cosserat-Rod-Kinematics-2020}}
\begin{equation}
  p_{\tmmathbf{a}}^{\tmmathbf{b}} = \sqrt{\frac{1 +\tmmathbf{a} \cdot
  \tmmathbf{b}}{2}} + \frac{1}{2} \tmmult \frac{\tmmathbf{a} \times
  \tmmathbf{b}}{\sqrt{\frac{1 +\tmmathbf{a} \cdot \tmmathbf{b}}{2}}} .
  \label{eq: paralleltransport}
\end{equation}
The proof is as follows. First it can be verified that
$p_{\tmmathbf{a}}^{\tmmathbf{b}}$ is a unit quaternion, as can be shown by
using the identity $\frac{|\tmmathbf{a} \times \tmmathbf{b}|^2}{1
+\tmmathbf{a} \cdot \tmmathbf{b}} = \frac{1 - (\tmmathbf{a} \cdot
\tmmathbf{b})^2}{1 +\tmmathbf{a} \cdot \tmmathbf{b}} = 1 -\tmmathbf{a} \cdot
\tmmathbf{b}$. Second, the rotation $p_{\tmmathbf{a}}^{\tmmathbf{b}}$ indeed
maps $\tmmathbf{a}$ to
\begin{equation}
  p_{\tmmathbf{a}}^{\tmmathbf{b}} \ast \tmmathbf{a}=
  p_{\tmmathbf{a}}^{\tmmathbf{b}} \tmmult \tmmathbf{a} \tmmult
  \overline{p_{\tmmathbf{a}}^{\tmmathbf{b}}} =\tmmathbf{b},
  \label{eq:parallel-transport-a-to-b}
\end{equation}
as can be checked by explicit calculation. Finally, the axis of $p_{\tmmathbf{a}}^{\tmmathbf{b}}$ is
indeed about the binormal $\tmmathbf{a} \times \tmmathbf{b}$:
equation~(\ref{eq:unit-quaternion-normal-form}) shows that the vector part of
the unit quaternion is aligned with the rotation axis and equation~(\ref{eq:
paralleltransport}) shows that the vector part of
$p_{\tmmathbf{a}}^{\tmmathbf{b}}$ is aligned with $\tmmathbf{a} \times
\tmmathbf{b}$.

For two units vectors $\tmmathbf{a}$ and $\tmmathbf{b}$ such that
$\tmmathbf{a}= -\tmmathbf{b}$, the parallel transport
$p_{\tmmathbf{a}}^{\tmmathbf{b}}$ is ill-defined.

\subsection{Reference and current configurations}\label{ssec:reference-vs-current}

\begin{figure}
  \centerline{\includegraphics[width=.99\textwidth]{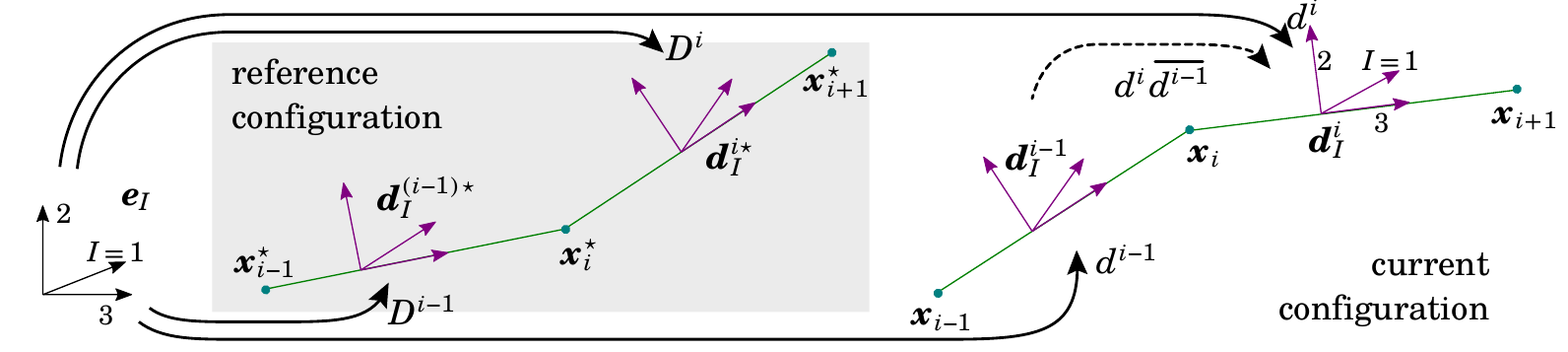}}
  \caption{A node $\tmmathbf{x}_i$, its adjacent segments, and the adjacent
  nodes $\tmmathbf{x}_{i \pm 1}$ in reference (gray background) and current
  (white background) configurations. Director frames, shown in purple, are
  represented by a unit quaternion, whose action on the Cartesian frame
  $\tmmathbf{e}_I$ yields the director frame.
  \label{fig:transportloop}}
\end{figure}

A configuration of the discrete rod is defined by a set of nodes
$\tmmathbf{x}_i$ indexed by an integer $i$, $0 \leqslant i \leqslant N$. We
consider an open rod having unconstrained endpoints $\tmmathbf{x}_0$ and
$\tmmathbf{x}_N$ for the moment; alternate boundary conditions such as
periodic or clamped boundary conditions are discussed later. For simplicity,
we limit attention to the case where the nodes are equally spaced in the
undeformed configuration, {\tmem{i.e.}}, the undeformed length $\ell^j$ is
independent of the segment index $j$: it is denoted as
\[ \ell^j = \ell . \]
In addition to the undeformed configuration, the simulation deals with two
configurations shown in Figure~\ref{fig:transportloop}:
\begin{itemize}
  \item {\bf Reference configuration} (shown with a gray background in the
  figure).
The only role of the reference configuration is to allow a parameterization of
the current configuration. It does not bear any physical meaning and 
its choice does not affect the results of the simulations.  It is chosen
for convenience.

In the reference configuration, the position of node $i$ is denoted by
$\tmmathbf{x}_i^{\star}$. The orthonormal frame of directors on segment $i$
connecting nodes $\tmmathbf{x}_i^{\star}$ and $\tmmathbf{x}_{i + 1}^{\star}$
is denoted as $(\tmmathbf{d}_I^{i \star})_{I \in \{ 1, 2, 3 \}}$. The
adaptation condition from equation~(\ref{eq:adaptation-continuous}) requires
that the third director $\tmmathbf{d}_3^{j \star}$ coincides with the unit
tangent $\tmmathbf{T}^j$ to the segment in reference configuration,
\begin{equation}
  \tmmathbf{d}_3^{j \star} =\tmmathbf{T}^j, \text{ where $\tmmathbf{T}^j =
  \frac{\tmmathbf{x}_{j + 1}^{\star} -\tmmathbf{x}_j^{\star}}{|
  \tmmathbf{x}_{j + 1}^{\star} -\tmmathbf{x}_j^{\star} |}$} .
  \label{eq:adaptation-constraint-reference}
\end{equation}

  \item {\bf Current configuration} (shown with a white background).
  The current configuration is the physical configuration of the rod and 
  is the unknown in a simulation.  It is parameterized by the
  degrees of freedoms (see Section~\ref{sec:discrete-kinematics}\ref{ssec:discrete-kinematics-summary}).
   
   In the current configuration, the center-line of the rod is defined by
the node positions $\tmmathbf{x}_i$. On segment $i$ connecting the nodes
$\tmmathbf{x}_i$ and $\tmmathbf{x}_{i + 1}$, the directors are denoted as
$(\tmmathbf{d}_I^i)_{I \in \{ 1, 2, 3 \}}$. The adaptation condition from
equation~(\ref{eq:adaptation-continuous}) requires
\begin{equation}
  \tmmathbf{d}_3^j =\tmmathbf{t}^j, \text{ where $\tmmathbf{t}^j =
  \frac{\tmmathbf{x}_{j + 1} -\tmmathbf{x}_j}{| \tmmathbf{x}_{j + 1}
  -\tmmathbf{x}_j |}$} . \label{eq:adaptation-constraint-current}
\end{equation}

\end{itemize}

As shown in the figure, the orthonormal director frames $(\tmmathbf{d}_I^{j
\star})_{1 \leqslant I \leqslant 3}$ and $(\tmmathbf{d}_I^j)_{1 \leqslant I
\leqslant 3}$ are represented by unit quaternions $D^j$ and $d^j$,
respectively, that yield the directors when applied to the Cartesian basis
$\tmmathbf{e}_I$:
\begin{equation}
  D^j \ast \tmmathbf{e}_I =\tmmathbf{d}^{j \star}_I \qquad d^j \ast
  \tmmathbf{e}_I =\tmmathbf{d}^j_I \text{\qquad for $I = 1, 2, 3$}.
  \label{eq:quaternion-representation-of-directors}
\end{equation}
The quaternions $d^{j \star}$ and $d^j$ therefore represent the
rotations $\sum_{I = 1}^3 \tmmathbf{d}_I^{j \star} \otimes
\tmmathbf{e}_I$ and $\sum_{I = 1}^3 \tmmathbf{d}_I^j \otimes
\tmmathbf{e}_I$, respectively.  They fully
describe their respective frames.

The reference and current configurations are not assumed to be close
to one another.  However, our parameterization introduces a weak
restriction: the reference configuration must be chosen such that the
angle of the rotation $\left( d^j \tmmult \overline{D}^j \right)$
mapping $\tmmathbf{d}_I^{j \star}$ to \ $\tmmathbf{d}_I^j$ does not
come close to $\mathpi$, in any of the segments $j$.
This condition is fulfilled by resetting
periodically the reference configuration to the current configuration:
\begin{itemize}
  \item in dynamic simulations, this reset is typically done at the end of any
  time step;
  
  \item in equilibrium problems, it is typically done whenever an equilibrium
  has been found and the load is incremented.
\end{itemize}
In principle, it is even possible to reset the reference configuration in the
middle of the Newton-Raphson iteration used to update a time step (in the dynamic case)
or the non-linear equilibrium (in the static case), but special care is
required as this amounts to changing the parameterization of the unknown during
iteration.

All the applications shown at the end of this paper deal with the static case,
{\tmem{i.e.}}, they involve the calculation of equilibria for a series of load
values: our simulations are initialized with the reference configuration
$\tmmathbf{x}_i^{\star}$, $\tmmathbf{d}^{j \star}_I$ representing a simple
starting point which is typically a straight or circular equilibrium
configuration without any load (see the example description for
further details). The reference configuration is reset each time an
equilibrium is found.

\subsection{Centerline-twist representation}\label{ssec:centerline-twist}

In this section, we introduce a parameterization that provides a concise representation 
of the current configuration
that is at the heart of the Discrete elastic rod method. 
All quantities from the reference configuration, such as the node positions
$\tmmathbf{x}_i^{\star}$, unit tangents $\tmmathbf{T}^j$, material frames
$\tmmathbf{d}_3^{j \star}$ and associated rotations $D^j$, are known. We
proceed to analyze the current configuration. A key observation is that
equation~(\ref{eq:adaptation-constraint-current}) yields the tangent director
$\tmmathbf{d}_3^j$ as a function of the node positions $\tmmathbf{x}_i$: if
the nodes are prescribed, the full frame of directors $\tmmathbf{d}_I^j$ can
only twist about this tangent. The three directors $(\tmmathbf{d}_I^j)_{1
\leqslant I \leqslant 3}$ on segment $j$, as well as the associated unit
quaternion $d^j$ by
equation~(\ref{eq:quaternion-representation-of-directors}), can therefore be
parameterized in terms of
\begin{itemize}
  \item the adjacent nodes positions $\tmmathbf{x}_j$ and $\tmmathbf{x}_{j +
  1}$,
  
  \item a scalar twist angle $\varphi^j$.
\end{itemize}
The parameterization used by the Discrete elastic rod method
may be written as~{\cite{Bergou-Wardetzky-EtAl-Discrete-Elastic-Rods-2008,Bergou-Audoly-EtAl-Discrete-Viscous-Threads-2010,Audoly-Clauvelin-EtAl-A-discrete-geometric-approach-2013}}
\begin{equation}
  d^j (\tmmathbf{x}_j, \varphi^j, \tmmathbf{x}_{j + 1}) = p^j (\tmmathbf{x}_j,
  \tmmathbf{x}_{j + 1}) \tmmult r_{\tmmathbf{T}^j} (\varphi^j) \tmmult D^j,
  \label{eq:centerline-twist-frame-reconstruction}
\end{equation}
where $\tmmathbf{x}_j$ and $\tmmathbf{x}_{j + 1}$ are the positions of the
adjacent nodes, $\varphi^j$ is the twisting angle,
\begin{equation}
  p^j (\tmmathbf{x}_j, \tmmathbf{x}_{j + 1}) =
  p_{\tmmathbf{T}^i}^{\tmmathbf{t}^i (\tmmathbf{x}_j, \tmmathbf{x}_{j + 1})}
  \label{eq:parallel-transport-in-time}
\end{equation}
is the parallel transport from the reference unit tangent $\tmmathbf{T}^i$ to
the current unit tangent $\tmmathbf{t}^i (\tmmathbf{x}_j, \tmmathbf{x}_{j +
1})$ given as a function of the node positions by
equation~(\ref{eq:adaptation-constraint-current}), $r_{\tmmathbf{T}^j}
(\varphi^j) = \cos \frac{\varphi^j}{2} +\tmmathbf{T}^j \tmmult \sin
\frac{\varphi^j}{2}$ is the rotation about $\tmmathbf{T}^j$ with angle
$\varphi^j$ (see equation~(\ref{eq:unit-quaternion-normal-form})), and $D^j$ is
the unit quaternion associated with the reference configuration of the
directors (see equation~(\ref{eq:quaternion-representation-of-directors})).

Using equations~(\ref{eq:quaternion-representation-of-directors}),
(\ref{eq:adaptation-constraint-reference})
and~(\ref{eq:parallel-transport-a-to-b}), we have $\tmmathbf{d}^j_3 =
d^j (\tmmathbf{x}_j, \varphi^j, \tmmathbf{x}_{j + 1}) \ast
\tmmathbf{e}_3 = p^j (\tmmathbf{x}_j, \tmmathbf{x}_{j + 1}) \ast
(r_{\tmmathbf{T}} (\varphi^j, \tmmathbf{T}^j) \ast (D^j \ast
\tmmathbf{e}_3)) = p_{\tmmathbf{T}^i}^{\tmmathbf{t}^i} \ast
(r_{\tmmathbf{T}^j} (\varphi^j) \ast \tmmathbf{T}^j) =
p_{\tmmathbf{T}^i}^{\tmmathbf{t}^i} \ast \tmmathbf{T}^j
=\tmmathbf{t}^j$: the
parameterization~(\ref{eq:centerline-twist-frame-reconstruction}) of
the directors satisfies the adaptation constraint
in~(\ref{eq:adaptation-constraint-current}) automatically.

This yields a parameterization of the rod in terms of the degrees of freedom
vector
\begin{equation}
  \tmmathbf{X}= (\tmmathbf{x}_0, \varphi^0, \tmmathbf{x}_1, \varphi^1,
  \tmmathbf{x}^2, \cdots, \tmmathbf{x}_{n - 1}, \varphi^{n - 1},
  \tmmathbf{x}_n), \label{eq:dofs}
\end{equation}
where the nodes positions $\tmmathbf{x}_i$ are read off directly from
$\tmmathbf{X}$ and the directors are reconstructed using
equations~(\ref{eq:quaternion-representation-of-directors})
and~(\ref{eq:centerline-twist-frame-reconstruction}). It is called the
{\tmem{centerline-twist representation}}.

As observed in Section~\ref{sec:discrete-kinematics}\ref{ssec:parallel_transport}, the parallel transport
in equation~(\ref{eq:parallel-transport-in-time}) is singular if
$\tmmathbf{t}^i (\tmmathbf{x}_j, \tmmathbf{x}_{j + 1}) = -\tmmathbf{T}^i$,
{\tmem{i.e.}}, if any one of the tangents flips by an angle $\mathpi$ between
the reference and current configuration. The periodic reset of the reference
configuration described earlier in Section~\ref{sec:discrete-kinematics}\ref{ssec:reference-vs-current}
prevents this from happening.

Note that in the original paper of
{\cite{Bergou-Wardetzky-EtAl-Discrete-Elastic-Rods-2008}}, parallel transport
was used to move the directors from one segment to an adjacent segment
({\tmem{spatial}} parallel transport). This makes the directors dependent on
the degrees of freedom associated with all the nodes and segments located on
one side of the directors. Here, like in subsequent work by the same
authors~{\cite{Bergou-Audoly-EtAl-Discrete-Viscous-Threads-2010,Audoly-Clauvelin-EtAl-A-discrete-geometric-approach-2013}},
we use parallel transport `in time': in
equation~(\ref{eq:centerline-twist-frame-reconstruction}), $p^j
(\tmmathbf{x}_j, \tmmathbf{x}_{j + 1})$ serves to parameterize the directors
in current configuration in terms of the same set of directors in reference
configuration. With this approach, the directors are a function of the
{\tmem{local}} degrees of freedom, as implied by the notation $d^j
(\tmmathbf{x}_j, \varphi^j, \tmmathbf{x}_{j + 1})$ in
equation~(\ref{eq:centerline-twist-frame-reconstruction}).

\subsection{Lagrangian rotation
gradient}\label{ssec:Lagrangian-rotation-gradient}

The rotation mapping one director frame $(\tmmathbf{d}_I^{i - 1})_{I = 1, 2,
3}$ to the adjacent director frame $(\tmmathbf{d}_I^i)_{I = 1, 2, 3}$ is shown
by the dashed arrow on top of Figure~\ref{fig:transportloop}. It captures the
variation of the frame along the rod, and it is the discrete counterpart of
the rotation gradient $\tmmathbf{\kappa} (s)$ introduced in
equation~(\ref{eq:rotation-gradient-gradient}). Using
equation~(\ref{eq:quaternion-representation-of-directors}), it can be written
as the composition of the rotations \ $\overline{d^{i - 1}}$ and $d^i$:
\[ d^i \tmmult \overline{d^{i - 1}} \of \tmmathbf{d}_I^{i - 1} \mapsto
   \tmmathbf{d}_I^i . \]
This rotation is an Eulerian quantity: like its continuous counterpart
$\tmmathbf{\kappa} (S)$, it is not invariant when the rod rotates
rigidly.  The following, however, is a Lagrangian version $q_i$ of the
rotation gradient that is invariant by rigid-body rotations,
\begin{equation}
  q_i (\tmmathbf{x}_{i - 1}, \varphi^{i - 1}, \tmmathbf{x}_i, \varphi^i,
  \tmmathbf{x}_{i + 1}) := \overline{d^{i - 1}} (\tmmathbf{x}_{i - 1},
  \varphi^{i - 1}, \tmmathbf{x}_i) \tmmult d^i (\tmmathbf{x}_i, \varphi^i,
  \tmmathbf{x}_{i + 1}) . \label{eq:rotation-gradient}
\end{equation}
Here, we depart from earlier work on Discrete elastic
rods~{\cite{Bergou-Wardetzky-EtAl-Discrete-Elastic-Rods-2008}} who
used $q_i := q_i^{\text{avg}} = \bar{d}_i^{\dag} \tmmult (d^i \tmmult
\overline{d^{i - 1}}) \tmmult d_i^{\dag}$ instead, where $d_i^{\dag}$
is some average of the adjacent frames $d^{i - 1}$ and $d^i$.  A
definition of the rotation gradient similar
to~(\ref{eq:rotation-gradient}) has been used in the context of
shearable
rods~{\cite{Gazzola-Dudte-EtAl-Forward-and-inverse-problems-2018}} and
in a purely geometric analysis of discrete
rods~{\cite{Linn-Discrete-Cosserat-Rod-Kinematics-2020}}.

We now explain why this definition represents a Lagrangian rotation
gradient.  One way to define a Lagrangian rotation gradient, is to
pull back the Eulerian rotation gradient $d^i \tmmult \overline{d^{i -
1}}$ to the reference configuration.  However, the discreteness of our
representation raises a difficulty: the frames are defined on the
segment while the Eulerian rotation gradient $d^i \tmmult
\overline{d^{i - 1}}$ is defined on the nodes.  So, we could use the
frame associated with the segment on the left of the node for the pull
back by defining $q_i^{\text{left}} = \overline{d^{i - 1}} \tmmult
(d^i \tmmult \overline{d^{i - 1}}) \tmmult d^{i - 1}$, but this biases
the choice on the left.  Or, we could use the right counter-part,
$q_i^{\text{right}} = \overline{d^i} \tmmult (d^i \tmmult
\overline{d^{i - 1}}) \tmmult d^i$, but this biases the choice to the
right.  However, these biases are apparent only: elementary
calculations shows that these are in fact identical
\begin{equation}
q_i^{\text{left}} = \overline{d^{i - 1}} \tmmult d^i \tmmult \left(
\overline{d^{i - 1}} \tmmult d^{i - 1} \right) = \overline{d^{i - 1}} \tmmult
d^i = q_i, \quad 
q_i^{\text{right}} = \left( \overline{d^i} \tmmult d^i \right)
\tmmult \overline{d^{i - 1}} \tmmult d^i = \overline{d^{i - 1}} \tmmult d^i
= q_i,
\end{equation}
thereby justifying our definition.

The unit quaternion $q_i$ introduced in equation~(\ref{eq:rotation-gradient})
is the discrete analogue of the pull-back $(\tmmathbf{e}_I \otimes
\tmmathbf{d}_I (s)) \cdot \tmmathbf{\kappa} (s)$ of the rotation gradient
$\tmmathbf{\kappa} (s)$ used in the continuous rod theory, whose components
$\kappa_J (s) =\tmmathbf{e}_J \cdot [(\tmmathbf{e}_I \otimes \tmmathbf{d}_I
(s)) \cdot \tmmathbf{\kappa} (s)] =\tmmathbf{d}_J (s) \cdot \tmmathbf{\kappa}
(s)$ define the bending and twisting measures. In the following section, bending
and twisting are similarly extracted from the unit quaternion $q_i$.

\subsection{Bending and twisting deformation measures}\label{ssec:strmeasures}

The discrete bending and twisting deformation measures are defined as the components of the
pure vector,
\begin{equation}
  \tmmathbf{\kappa}_i (\tmmathbf{x}_{i - 1}, \varphi^{i - 1}, \tmmathbf{x}_i,
  \varphi^i, \tmmathbf{x}_{i + 1}) = q_i - \overline{q}_i . \label{eq:kappi}
\end{equation}
This $\tmmathbf{\kappa}_i$ is twice the vector part $\mathcal{I} (q_i) =
\frac{q_i - \overline{q}_i}{2}$ of the quaternion $q_i$, which shows that it
is indeed a vector. Let $\kappa_{i, I}$ denote its components in the Cartesian
basis, such that $\tmmathbf{\kappa}_i = \sum_{I = 1}^3 \kappa_{i, I} \tmmult
\tmmathbf{e}_I$. The first two components $\kappa_{i, 1}$ and $\kappa_{i, 2}$
can be interpreted as measures of bending about the transverse directors
$\tmmathbf{d}_1^j$ and $\tmmathbf{d}_2^j$, while the third component
$\kappa_{i, 3}$ is a discrete measure of twisting. Like $q_i$, these are
{\tmem{integrated}} versions of their smooth counterparts, that are
proportional to the discretization length $\ell$; this will be taken into
account when setting up a discrete strain energy.

\subsection{Summary}\label{ssec:discrete-kinematics-summary}

The current configuration is reconstructed in terms of the degrees of freedom
$\tmmathbf{X}$ from equation~(\ref{eq:dofs}) as follows:
\begin{itemize}
  \item the node positions $\tmmathbf{x}_i$ are directly extracted from
  $\tmmathbf{X}$, see equation~(\ref{eq:dofs}),
  
  \item the unit tangents $\tmmathbf{t}^j (\tmmathbf{x}_j, \tmmathbf{x}_{j +
  1})$ are obtained from equation~(\ref{eq:adaptation-constraint-current}),
  
  \item parallel transport $p^j (\tmmathbf{x}_j, \tmmathbf{x}_{j + 1})$ is
  obtained by combining equations~(\ref{eq:parallel-transport-in-time})
  and~(\ref{eq: paralleltransport}),
  
  \item the director frames $d^j (\tmmathbf{x}_j, \varphi^j, \tmmathbf{x}_{j +
  1})$ are obtained from
  equation~(\ref{eq:centerline-twist-frame-reconstruction}),
  
  \item the rotation gradient $q_i (\tmmathbf{x}_{i - 1}, \varphi^{i - 1},
  \tmmathbf{x}_i, \varphi^i, \tmmathbf{x}_{i + 1})$ is available from
  equation~(\ref{eq:rotation-gradient}),
  
  \item the bending and twisting deformation vector $\tmmathbf{\kappa}_i
  (\tmmathbf{x}_{i - 1}, \varphi^{i - 1}, \tmmathbf{x}_i, \varphi^i,
  \tmmathbf{x}_{i + 1})$ is calculated from equation~(\ref{eq:kappi}).
\end{itemize}

Finally, a possible definition of the discrete stretching measure on segment
$j$ joining nodes $\tmmathbf{x}_j$ and $\tmmathbf{x}_{j + 1}$ is
\begin{equation}
  \varepsilon^j (\tmmathbf{x}_j, \tmmathbf{x}_{j + 1}) = \frac{1}{2} \tmmult
  \left( \frac{(\tmmathbf{x}_{j + 1} -\tmmathbf{x}_j)^2}{\ell} - \ell \right),
  \label{eq:stretching-strain}
\end{equation}
see for
instance~{\cite{Lestringant-Audoly-EtAl-A-discrete-geometrically-exact-2020}}.
Here, $\ell$ denotes the undeformed length of the segments, which is different
from the length $| \tmmathbf{x}_{j + 1}^{\star} -\tmmathbf{x}_j^{\star} |$ in
reference configuration. This discrete stretching measure is an integrated
version of the continuous strain $\varepsilon (S)$, like the discrete bending
and twisting deformation measures $\kappa_{i, I}$. The particular definition of the
stretching measure $\varepsilon^j$ in equation~(\ref{eq:stretching-strain}) requires the
evaluation of the {\tmem{squared}} norm and not of the norm itself, which
simplifies the calculation of the gradient significantly.

\subsection{Interpretation of the discrete
deformation measures}\label{ssec:discrete-strain-interpretation}

We now show that the discrete deformation measures (up to a minor rescaling) 
may be interpreted as the rotation that transports the director frame from one 
segment to the next.

Consider the  function $\psi$ 
\begin{equation}
  \psi (t) = \frac{\arcsin  (t / 2)}{t / 2} \text{\quad for $0 \leqslant t
  \leqslant 2$}, \label{eq:psi}
\end{equation}
and note that $\psi(t) \approx 1$ for $t \ll 1$ (See supplementary information for 
a plot of this function).  Define the {\tmem{adjusted deformation measure}} to be
\begin{equation}
  \omega_{i, J} = \psi (| \tmmathbf{\kappa}_i |) \tmmult \tmmathbf{\kappa}_i
  \cdot \tmmathbf{e}_J . \label{eq:from-kappai-to-omegai}
\end{equation}
This is well defined for all values of $ \kappa$ since 
$| \tmmathbf{\kappa}_i | = | q_i - \overline{q}_i |
\leqslant 2 \tmmult | q_i | = 2$.
This rescaling is insignificant in the continuum limit where $d^{i - 1}
\approx d^i$, $q_i \approx 1$ and $| \tmmathbf{\kappa}_i | \ll 1$, implying
$\psi (| \tmmathbf{\kappa}_i |) \approx 1$. Even for moderate values of $|
\tmmathbf{\kappa}_i |$, the original and adjusted deformations measures are not very
different, $\omega_{i, J} \approx \tmmathbf{\kappa}_i \cdot \tmmathbf{e}_J$,
as the variations of the function $\psi$ are bounded by $1 \leqslant \psi (t)
\leqslant \pi/2 $.

The adjusted deformation measure has a simple geometric interpretation.
We start from the decomposition~(\ref{eq:unit-quaternion-normal-form}) of the
rotation gradient $q_i = r_{\tmmathbf{n}_i} (\theta_i) = \cos
\frac{\theta_i}{2} +\tmmathbf{n}_i \tmmult \sin \frac{\theta_i}{2} = \exp
\frac{\tmmathbf{n}_i \tmmult \theta_i}{2}$, where $\tmmathbf{n}_i$ is a unit
vector aligned with the axis of the rotation $q_i$, and $\theta_i$ is the
angle of this rotation, $0 \leqslant \theta \leqslant \mathpi$. In view of
equation~(\ref{eq:kappi}), $\tmmathbf{\kappa}_i = q_i - \overline{q}_i = 2
\tmmult \sin \frac{\theta_i}{2} \tmmult \tmmathbf{n}_i$. In particular, $|
\tmmathbf{\kappa}_i | = 2 \tmmult \sin \frac{\theta_i}{2}$ and so $\psi (|
\tmmathbf{\kappa}_i |) = \frac{\theta_i / 2}{\sin (\theta_i / 2)}$ from
equation~(\ref{eq:psi}). The adjusted strain is then $\omega_{i, J} \tmmult
\tmmathbf{e}_J = \psi (| \tmmathbf{\kappa}_i |) \tmmult \tmmathbf{\kappa}_i =
\frac{\theta_i / 2}{\sin (\theta_i / 2)} \tmmult 2 \tmmult \sin
\frac{\theta_i}{2} \tmmult \tmmathbf{n}_i = \theta_i \tmmult \tmmathbf{n}_i$:
in effect, the adjustment factor $\psi (| \tmmathbf{\kappa}_i |)$ transforms
$\tmmathbf{\kappa}_i = 2 \tmmult \mathcal{I} (q_i)$ (twice the vector part of
$q_i$) into $\omega_{i, J} \tmmult \tmmathbf{e}_J = \theta_i \tmmult
\tmmathbf{n}_i = 2 \tmmult \log q_i$ (twice its logarithm).

Now, rewriting $q_i = \overline{d^{i - 1}} \tmmult d^i = \overline{d^{i - 1}}
\tmmult \left( d^i \tmmult \overline{d^{i - 1}} \right) \tmmult d^{i - 1} =
q^{\text{right}}_i$, one sees that $q_i$ is conjugate to $d^i \tmmult
\overline{d^{i - 1}}$. Combining with $q_i = \cos \frac{\theta_i}{2}
+\tmmathbf{n}_i \tmmult \sin \frac{\theta_i}{2}$, we have $d^i \tmmult
\overline{d^{i - 1}} = d^{i - 1} \tmmult q_i \tmmult \overline{d^{i - 1}} =
\cos \frac{\theta_i}{2} + (d^{i - 1} \ast \tmmathbf{n}_i) \tmmult \sin
\frac{\theta_i}{2} = \exp \frac{(d^{i - 1} \ast \tmmathbf{n}_i) \tmmult
\theta_i}{2}$: as is well known, the conjugate rotation $d^i \tmmult
\overline{d^{i - 1}}$ has the same angle $\theta_i$ as the original rotation
$q_i $ and its axis is obtained by applying the rotation $d^{i - 1}$ to the
original axis. This can be rewritten as
\begin{equation}
  d^i = \exp \left( \frac{\tmmathbf{\Omega}_i}{2} \right) \tmmult d^{i - 1}
  \label{eq:from-left-to-right-frame}
\end{equation}
where $\tmmathbf{\Omega}_i = d^{i - 1} \ast \tmmathbf{n}_i \tmmult \theta_i =
d^{i - 1} \ast \omega_{i, J} \tmmult \tmmathbf{e}_J = \omega_{i, J} \tmmult
\tmmathbf{d}_J^{i - 1}$ is a (finite) rotation vector. Similar relations have
been derived in the work
of~{\cite{Linn-Discrete-Cosserat-Rod-Kinematics-2020}}.
Repeating the same argument with $q_i = \overline{d^{i - 1}} \tmmult d^i =
\overline{d^i} \tmmult (d^i \tmmult \overline{d^{i - 1}}) \tmmult d^i =
q^{\text{left}}_i$, \ one can show that the vector $\tmmathbf{\Omega}$ has the
same decomposition in the other directors frame, $\tmmathbf{\Omega}_i =
\omega_{i, J} \tmmult \tmmathbf{d}_J^i$:
\begin{equation}
  \tmmathbf{\Omega}_i = \omega_{i, J} \tmmult \tmmathbf{d}_J^{i - 1} =
  \omega_{i, J} \tmmult \tmmathbf{d}_J^i .
  \label{eq:rotation-vector-from-left-to-right-frame}
\end{equation}
Equations~(\ref{eq:from-left-to-right-frame}--\ref{eq:rotation-vector-from-left-to-right-frame})
show that {\tmem{the adjusted deformation measures $\omega_{i, J}$ are the
components of the rotation vector $\tmmathbf{\Omega}_i$ that maps one set of
directors frame $(\tmmathbf{d}_I^{i - 1})_{I = 1, 2, 3}$ to the other one
$(\tmmathbf{d}_I^i)_{I = 1, 2, 3}$}} across the vertex $\tmmathbf{x}_i$.
Remarkably, these components can be calculated in any one of the 
adjacent directors frame as they are identical.

One could build a Discrete elastic rod model based on the adjusted
deformation measure $\omega_{i, J} \tmmult \tmmathbf{e}_J = 2 \tmmult
\overline{d^{i - 1}} \ast \log \left( d^i \tmmult \overline{d^{i - 1}}
\right) = 2 \tmmult \overline{d^i} \ast \log \left( d^i \tmmult
\overline{d^{i - 1}} \right)$ instead of the deformation measure
$\tmmathbf{\kappa}_i$ proposed in
Section~\ref{sec:discrete-kinematics}\ref{ssec:strmeasures}.  The
benefit is that $\omega_{i, J}$ have an even simpler interpretation,
see
equations~(\ref{eq:from-left-to-right-frame}--\ref{eq:rotation-vector-from-left-to-right-frame}).
The drawback is that the function $\psi$ gets involved in the
calculation of the strain, resulting in cumbersome formulas for the
strain gradients (Section \ref{sec:variations}).  Therefore, we
continue to use the original deformation measures.

\section{Variations of the discrete deformation measures}\label{sec:variations}

In this section, we present explicit formulae for the first and second
derivatives of the deformation measures $\tmmathbf{\kappa}_i$
(summarized in
Section~\ref{sec:discrete-kinematics}\ref{ssec:discrete-kinematics-summary})
with respect to $\tmmathbf{X}$.  The first gradient is required for
determination of the internal forces, which are the first gradient of
the strain energy.  The availability of the second gradient in
analytical form makes it possible to use implicit time-stepping
methods (in dynamic problems) or to evaluate the Hessian for second
order methods (in static problems).

Our notation for variations is first introduced based on a simple example. For
a function $\tmmathbf{y}=\tmmathbf{f} (\tmmathbf{x})$ taking a vector argument
$\tmmathbf{x}$ and returning a vector $\tmmathbf{y}$, the first variation is
the linear mapping $\delta \tmmathbf{x} \mapsto \delta
\tmmathbf{y}=\tmmathbf{f}' (\tmmathbf{x}) \cdot \delta \tmmathbf{x}$, where
$\delta \tmmathbf{x}$ is a perturbation to $\tmmathbf{x}$ and $\tmmathbf{f}'
(\tmmathbf{x})$ is the gradient matrix. To compute the second variation, we
start from $\delta \tmmathbf{y}=\tmmathbf{f}' (\tmmathbf{x}) \cdot \delta
\tmmathbf{x}$, perturb the argument $\tmmathbf{x}$ of $\tmmathbf{f}'$ as
$\tmmathbf{x}+ \delta \tmmathbf{x}$ and linearize the result as $\tmmathbf{f}'
(\tmmathbf{x}+ \delta \tmmathbf{x}) \cdot \delta \tmmathbf{x} \approx
\tmmathbf{f}' (\tmmathbf{x}) \cdot \delta \tmmathbf{x}+\tmmathbf{f}''
(\tmmathbf{x}) \of (\delta \tmmathbf{x} \otimes \delta \tmmathbf{x})$. Here,
the second variation is defined as the second order term $\delta^2
\tmmathbf{y} \assign \tmmathbf{f}'' (\tmmathbf{x}) \of (\delta \tmmathbf{x}
\otimes \delta \tmmathbf{x})$, where $\tmmathbf{f}'' (\tmmathbf{x})$ is the
Hessian. By construction, $\delta^2 \tmmathbf{y}$ is a quadratic form of
$\delta \tmmathbf{x}$.

In this section, the reference configuration is fixed and the degrees of
freedom are perturbed by $\delta \tmmathbf{X}= (\cdots, \delta \tmmathbf{x}_i,
\delta \varphi^i, \cdots)$. 
We simply present the final results; the detailed calculations are cumbersome but straightforward, and 
provided as supplementary material.

\begin{itemize}
  \item {\tmstrong{unit tangents}} $\tmmathbf{t}^i = (\tmmathbf{x}_{i + 1}
  -\tmmathbf{x}_i) / | \tmmathbf{x}_{i + 1} -\tmmathbf{x}_i |$ from
  equation~(\ref{eq:adaptation-constraint-current}),
  \begin{equation}
    \begin{array}{rll}
      \delta \tmmathbf{t}^i & = & \frac{\tmmathbf{I}-\tmmathbf{t}^i \otimes
      \tmmathbf{t}^i}{|\tmmathbf{x}_{i + 1} -\tmmathbf{x}_i |} \cdot (\delta
      \tmmathbf{x}_{i + 1} - \delta \tmmathbf{x}_i)\\
      \delta^2 \tmmathbf{t}^i & = & - \frac{\tmmathbf{\tau}^i +
      (\tmmathbf{\tau}^i)^{T (132)} + (\tmmathbf{\tau}^i)^{T
      (231)}}{|\tmmathbf{x}_{i + 1} -\tmmathbf{x}_i |^2} \of ((\delta
      \tmmathbf{x}_{i + 1} - \delta \tmmathbf{x}_i) \otimes (\delta
      \tmmathbf{x}_{i + 1} - \delta \tmmathbf{x}_i)),
    \end{array} \label{eq:variations-t}
  \end{equation}
  where $\tmmathbf{I}$ is the identity matrix, $\tmmathbf{\tau}^i$ is the
  third-order tensor $\tmmathbf{\tau}^i = (\tmmathbf{I}-\tmmathbf{t}^i \otimes
  \tmmathbf{t}^i) \otimes \tmmathbf{t}^i$, the colon denotes the double
  contraction of the {\tmem{last}} two indices of the rank-three tensor on the
  left-hand side. For any permutation $(n_1, n_2, n_3)$ of $(1, 2, 3)$, $T
  (n_1, n_2, n_3)$ denotes the generalized transpose of a rank-three 
  tensor $\tmmathbf{\mu}$ such
  that $\mu^{T (n_1 n_2 n_3)}_{i_1 i_2
  i_3} = \mu_{i_{n_1} i_{n_2} i_{n_3}}$;
  
  \item {\tmstrong{parallel transport}} \ $p^i =
  p_{\tmmathbf{T}^i}^{\tmmathbf{t}^i}$ from
  equations~(\ref{eq:parallel-transport-in-time}) and~(\ref{eq:
  paralleltransport}),
  \begin{equation}
    \begin{array}{rll}
      \delta \hat{\tmmathbf{p}}^i & = & \left( (\tmmathbf{t}^i)_{\times} -
      \frac{\tmmathbf{t}^i \otimes \tmmathbf{k}^i}{2} \right) \cdot \delta
      \tmmathbf{t}^i, \\
      \delta^2 \hat{\tmmathbf{p}}^i & = & \left( (\tmmathbf{t}^i)_{\times} -
      \frac{\tmmathbf{t}^i \otimes \tmmathbf{k}^i}{2} \right) \cdot \delta^2
      \tmmathbf{t}^i + \left( \delta \tmmathbf{t}^i \cdot \frac{\tmmathbf{k}^i
      \otimes \tmmathbf{T}^i +\tmmathbf{T}^i \otimes \tmmathbf{k}^i}{4 \tmmult
      (1 +\tmmathbf{T}^i \cdot \tmmathbf{t}^i)} \cdot \delta \tmmathbf{t}^i
      \right) \tmmult \tmmathbf{t}^i - (\delta \tmmathbf{t}^i \otimes \delta
      \tmmathbf{t}^i) \cdot \frac{\tmmathbf{k}^i}{2}
    \end{array} \label{eq:variations-p}
  \end{equation}
  where for any vector $\tmmathbf{a}$, $\tmmathbf{a}_{\times}$ is the linear
  operator
  \begin{equation}
    \tmmathbf{a}_{\times} \of \tmmathbf{u} \mapsto \tmmathbf{a} \times
    \tmmathbf{u} \label{eq:cross-operator}
  \end{equation}
  and $\tmmathbf{k}^i$ is the binormal defined by
  \begin{equation}
    \tmmathbf{k}^i = \frac{2 \tmmult \tmmathbf{T}^i \times \tmmathbf{t}^i}{1
    +\tmmathbf{T}^i \cdot \tmmathbf{t}^i} \ ; \label{eq:variations-ki-def}
  \end{equation}
  \item {\tmstrong{directors rotation}} $d^i$ from
  equation~(\ref{eq:centerline-twist-frame-reconstruction}),
  \begin{equation}
    \begin{array}{rll}
      \delta \hat{\tmmathbf{d}}^i & = & \delta \varphi^i \tmmathbf{t}^i +
      \delta \hat{\tmmathbf{p}}^i, \\
      \delta^2 \hat{\tmmathbf{d}}^i & = & \delta \varphi^i \tmmult \delta
      \tmmathbf{t}^i + \delta^2 \hat{\tmmathbf{p}}^i ;
    \end{array} \label{eq:variations-di}
  \end{equation}
  \item {\tmstrong{rotation gradient}} $q_i$ from
  equation~(\ref{eq:rotation-gradient}),
  \begin{equation}
    \begin{array}{rll}
      \delta \hat{\tmmathbf{q}}_i & = & \overline{d^{i - 1}} \ast (\delta
      \hat{\tmmathbf{d}}^i - \delta \hat{\tmmathbf{d}}^{i - 1}),\\
      \delta^2 \hat{\tmmathbf{q}}_i & = & \overline{d^{i - 1}} \ast (\delta^2
      \hat{\tmmathbf{d}}^i - \delta^2 \hat{\tmmathbf{d}}^{i - 1}) + \delta
      \hat{\tmmathbf{q}}_i \times (\overline{d^{i - 1}} \ast \delta
      \hat{\tmmathbf{d}}^{i - 1}) ;
    \end{array} \label{eq:variations-qi}
  \end{equation}
  \item {\tmstrong{discrete bending and twisting strain measure vector}}
  $\tmmathbf{\kappa}_i$ from equation~(\ref{eq:kappi}),
  \begin{equation}
    \begin{array}{rll}
      \delta \tmmathbf{\kappa}_i & = & \mathcal{I} \left( \delta
      \hat{\tmmathbf{q}}_i \tmmult q_i \right),\\
      \delta^2 \tmmathbf{\kappa}_i & = & \mathcal{I} \left( \Bigl( \delta^2
      \hat{\tmmathbf{q}}_i - \frac{\delta \hat{\tmmathbf{q}}_i \cdot \delta
      \hat{\tmmathbf{q}}_i}{2} \Bigr) \tmmult q_i \right);
    \end{array} \label{eq:variations-kappai}
  \end{equation}
  where $\mathcal{I} (q) = \frac{q - \overline{q}}{2}$ denotes the vector part
  of a quaternion $q$.
  
  \item {\tmstrong{stretching measure	}}~$\varepsilon^i$ from
  equation~(\ref{eq:stretching-strain}),
  \begin{equation}
    \begin{array}{rll}
      \delta \varepsilon^i & = & \frac{\tmmathbf{x}_{i + 1}
      -\tmmathbf{x}_i}{\ell} \cdot (\delta \tmmathbf{x}_{i + 1} - \delta
      \tmmathbf{x}_i),\\
      \delta^2 \varepsilon^i & = & \frac{1}{\ell} \tmmult (\delta
      \tmmathbf{x}_{i + 1} - \delta \tmmathbf{x}_i) \cdot (\delta
      \tmmathbf{x}_{i + 1} - \delta \tmmathbf{x}_i) .
    \end{array} \label{eq:variations-epsiloni}
  \end{equation}
\end{itemize}
In these formula, the first and second variations of the rotations $p^i$,
$d^i$ and $q_i$ are not captured by quaternions but by regular
{\tmem{vectors}}, bearing a hat, such as $\delta \hat{\tmmathbf{p}}^i$,
$\delta^2 \hat{\tmmathbf{p}}^i$, $\delta \hat{\tmmathbf{d}}^i$, etc.
Equations~(\ref{eq:variations-t}--\ref{eq:variations-epsiloni}) involve
standard calculations from Euclidean geometry: the more advanced quaternion
calculus is only required in the proof given in 
the supplementary materials.

Equations~(\ref{eq:variations-t}--\ref{eq:variations-epsiloni})
suffice to calculate the strain gradients.  They can be implemented
easily and efficiently using standard libraries for vector and matrix
algebra.  These formulas for the first and second gradient of strain
are considerably simpler than those applicable to the discrete strain
measures used in earlier work on Discrete elastic
rods~{\cite{Bergou-Wardetzky-EtAl-Discrete-Elastic-Rods-2008,Audoly-Clauvelin-EtAl-A-discrete-geometric-approach-2013,Panetta-Konakovic-Lukovic-EtAl-X-Shells:-A-New-Class-of-Deployable-2019,Lestringant-Audoly-EtAl-A-discrete-geometrically-exact-2020}}.

In equations~(\ref{eq:variations-t}--\ref{eq:variations-epsiloni}), the
perturbations to the degrees of freedom such as $\delta \tmmathbf{x}_i$ and
$\delta \varphi^i$ are dummy variables. The first-order variations such as
$\delta \tmmathbf{t}^i$, $\delta \hat{\tmmathbf{p}}^i$, must be represented
numerically as linear forms, by storing their coefficients as vectors.
Similarly, the second-order variations such as $\delta^2 \tmmathbf{t}^i$,
$\delta^2 \hat{\tmmathbf{p}}^i$, etc.~are represented as quadratic forms,
whose coefficients are stored as sparse symmetric matrices; the reader is referred to~{\cite{Lestringant-Audoly-EtAl-A-discrete-geometrically-exact-2020}} for further
details on this aspect of implementation. 
All these coefficients depend on the current configuration and must be updated
whenever the degrees of freedom $\tmmathbf{X}$ or the reference configuration
change.

These vectors and symmetric matrices should be stored at an appropriate place
in the data structure representing the Discrete elastic rod. The tensors
representing $\delta \tmmathbf{t}^i$, $\delta \hat{\tmmathbf{p}}^i$, $\delta^2
\hat{\tmmathbf{p}}^i$ and $\delta^2 \hat{\tmmathbf{d}}^i$ depend on the
perturbations $\delta \tmmathbf{x}_i$ and $\delta \tmmathbf{x}_{i + 1}$ to the
nodes adjacent to a given segment, and therefore best stored in the data
structure representing segments, which have access naturally to the degrees of
freedom of the adjacent nodes. The quantities $\delta \hat{\tmmathbf{d}}^i$
and $\delta^2 \hat{\tmmathbf{d}}^i$ make use of the twisting angle $\delta
\varphi^i$ in addition to the adjacent nodes $\delta \tmmathbf{x}_i$ and
$\delta \tmmathbf{x}_{i + 1}$, and should be stored in the data structure
representing the material frame attached to particular segment. The quantities
$\delta \hat{\tmmathbf{q}}_i$, $\delta \tmmathbf{\kappa}_i$, $\delta^2
\hat{\tmmathbf{q}}_i$ and $\delta^2 \tmmathbf{\kappa}_i$ are best stored in a
data structure representing an elastic hinge at a node, that depends on the
material frames at the adjacent segments.

\section{Constitutive models}\label{sec:ribbons}

The discrete kinematics from Sections~\ref{sec:discrete-kinematics}
and~\ref{sec:variations} can be combined with a variety of constitutive laws
to produce discrete numerical models for rods that are elastic, viscous,
visco-elastic, etc.: the procedure has been documented in previous work, and
it is similar to the general approach used in finite-element analysis. Elastic
problems are treated by introducing a strain energy function $U
(\tmmathbf{X})$, whose gradient with respect to $\tmmathbf{X}$ yields  the negative of the
discrete elastic
forces~{\cite{Bergou-Wardetzky-EtAl-Discrete-Elastic-Rods-2008,Lestringant-Audoly-EtAl-A-discrete-geometrically-exact-2020}}; while 
viscous problems are treated by introducing a discrete Rayleigh potential $U
(\tmmathbf{X}, \dot{\tmmathbf{X}})$ ,whose gradient with respect to velocities
$\dot{\tmmathbf{X}}$ yields discrete viscous
forces~{\cite{Bergou-Audoly-EtAl-Discrete-Viscous-Threads-2010,Brun-Ribe-EtAl-A-numerical-investigation-of-the-fluid-2012,Audoly-Clauvelin-EtAl-A-discrete-geometric-approach-2013}}.
More advanced constitutive models such as visco-elastic laws can be treated by
variational constitutive updates of a discrete potential that makes use of the
same discrete deformation
measures~{\cite{Lestringant-Audoly-EtAl-A-discrete-geometrically-exact-2020}}.
In~{\cite{Lestringant-Audoly-EtAl-A-discrete-geometrically-exact-2020}}, it
is emphasized that these different constitutive models can be
implemented {\tmem{independently}} of the geometric definition of discrete
deformation measure. Using this decoupled approach, it is straightforward to combine the
kinematic element proposed in the present work with constitutive element from
previous work.  We illustrate this with the
classical, linearly elastic rod in Section~\ref{sec:ribbons}\ref{ssec:linearly-elastic-rods}
(Kirchhoff rod model), and a discrete inextensible ribbon model in
Section~\ref{sec:ribbons}\ref{ssec:ribbons} (Wunderlich model). The latter is a novel
application of the Discrete elastic rod method.

\subsection{Elastic rods (Kirchhoff model)}\label{ssec:linearly-elastic-rods}

The classical, continuous theory of elastic rods uses a strain energy
functional $U [\tmmathbf{\kappa}] = \int_0^L E (\kappa_{(1)} (s), \kappa_{(2)}
(s), \kappa_{(3)} (s)) \tmmult \mathd s$, where $\kappa_{(I)} (s)
=\tmmathbf{\kappa} (s) \cdot \tmmathbf{d}_I (s)$ are the components of the
rotation gradient in the frame of directors, see
equation~(\ref{eq:k-sub-I-continuous}). For an inextensible, linearly elastic
rod made of a Hookean material with natural curvature $\kappa_{(0)}$, for
instance, the strain energy density is 
\begin{equation}
  E (\kappa_{(1)} (s), \kappa_{(2)} (s), \kappa_{(3)} (s)) = \frac{1}{2}
  \tmmult Y \tmmult I_1 \tmmult \kappa_{(1)}^2 + \frac{1}{2} \tmmult Y \tmmult
  I_2 \tmmult (\kappa_{(2)} - \kappa_{(0)})^2 + \frac{1}{2} \tmmult \mu
  \tmmult J \tmmult \kappa_{(3)}^2 \label{eq:strain-energy-linearly-elastic}
\end{equation}
where $Y$ and $\mu$ are the Young modulus and the shear modulus of the
material, $I_1$ and $I_2$ are the geometric moments of inertia of the
cross-section, and $J$ is the torsional constant.

In the discrete setting, we introduce a strain energy ${\sum_{\text{$i$}}} 
E_i (\tmmathbf{\kappa}_i)$ where the sum runs over all interior nodes $i$. The
strain energy assigned to an interior node $i$ is defined in terms of the
strain energy density as
\begin{equation}
  E_i (\tmmathbf{\kappa}_i) = \ell \tmmult E \left(
  \frac{\tmmathbf{\kappa}_i}{\ell} \right),
  \label{eq:continuous-to-discrete-Ei}
\end{equation}
(no implicit sum over $i$), where $\ell$ is the undeformed length of
the segments for a uniform mesh.  The factor $\ell$ in the argument of
$E$ takes care of the fact that $\tmmathbf{\kappa}_i$ is an integrated
quantity, {\tmem{i.e.}}, it is $\frac{\tmmathbf{\kappa}_i}{\ell} \cdot
\tmmathbf{e}_J$ and not just $\tmmathbf{\kappa}_i \cdot
\tmmathbf{e}_J$ that converges to the continuous strain $\kappa_{(J)}
(s)$; for a non-uniform grid, this $\ell$ would need to be replaced
with the Voronoi length associated with the interior vertex $i$ in
undeformed configuration.  The factor $\ell$ in factor of $E$ in
equation~(\ref{eq:continuous-to-discrete-Ei})
ensures that
the discrete sum ${\sum_{\text{$i$}}} E_i = {\sum_{\text{$i$}}} \ell
\tmmult E$ converges to the integral $\int_0^L E \tmmult \mathd s =
U$~\cite{Bergou-Wardetzky-EtAl-Discrete-Elastic-Rods-2008}.

Consider for instance an equilibrium problem with dead forces
$\tmmathbf{F}_i$ on the nodes: it is governed by the total potential energy
$\Phi (\tmmathbf{X})$ defined in terms of $\tmmathbf{X}= (\tmmathbf{x}_0,
\varphi_0, \ldots, \varphi_{N - 1}, \tmmathbf{x}_N)$ as
\begin{equation}
  \Phi (\tmmathbf{X}) = \sum_{i = 1}^{N - 1} E_i (\tmmathbf{\kappa}_i
  (\tmmathbf{x}_{i - 1}, \varphi^{i - 1}, \tmmathbf{x}_i, \varphi^i,
  \tmmathbf{x}_{i + 1})) - \sum_{i = 0}^N \tmmathbf{F}_i \cdot \tmmathbf{x}_i
  . \label{eq:discrete-strain-energy-density}
\end{equation}
This energy is minimized subject to the inextensibility constraints
\begin{equation}
  \forall i \in (0, N - 1) \qquad \varepsilon^j (\tmmathbf{x}_j,
  \tmmathbf{x}_{j + 1}) = 0.
  \label{eq:elastic-simulations-inextensibility-cstraint}
\end{equation}
In
equations~(\ref{eq:discrete-strain-energy-density}--\ref{eq:elastic-simulations-inextensibility-cstraint}),
the elastic deformation measures $\tmmathbf{\kappa}_i$ and $\varepsilon^j$ is reconstructed
in terms of the unknown $\tmmathbf{X}$ by the method described in
Section~\ref{sec:discrete-kinematics}, as expressed by the notation
$\tmmathbf{\kappa}_i (\tmmathbf{x}_{i - 1}, \varphi^{i - 1}, \tmmathbf{x}_i,
\varphi^i, \tmmathbf{x}_{i + 1})$ and $\varepsilon^j (\tmmathbf{x}_j,
\tmmathbf{x}_{j + 1})$.

In the case of dead forces, the first and second variations of the total
potential energy is derived as
\begin{equation}
  \begin{array}{rll}
    \delta \Phi & = & \sum_{i = 1}^{N - 2} \dfrac{\partial E_i}{\partial
    \tmmathbf{\kappa}_i} \cdot \delta \tmmathbf{\kappa}_i - \sum_{i = 0}^{N -
    1} \tmmathbf{F}_i \cdot \delta \tmmathbf{x}_i\\
    \delta^2 \Phi & = & \sum_{i = 1}^{N - 2} \left( \delta \tmmathbf{\kappa}_i
    \cdot \dfrac{\partial^2 E_i}{\partial \tmmathbf{\kappa}_i^2} \cdot \delta
    \tmmathbf{\kappa}_i + \dfrac{\partial E_i}{\partial \tmmathbf{\kappa}_i}
    \of \delta^2 \tmmathbf{\kappa}_i \right),
  \end{array} \label{eq:strain-energy-gradient-hessian}
\end{equation}
see for instance
{\cite{Lestringant-Audoly-EtAl-A-discrete-geometrically-exact-2020}}. Here,
$\frac{\partial E_i}{\partial \tmmathbf{\kappa}_i}$ and $\frac{\partial^2
E_i}{\partial \tmmathbf{\kappa}_i^2}$ are the internal stress and tangent
elastic stiffness produced by the elastic constitutive model $E_i
(\tmmathbf{\kappa}_i)$. The two terms appearing in the parentheses in the
right-hand side of $\delta^2 \Phi$ are known as the elastic and geometric
stiffness, respectively. The first and second variations of the strain,
$\delta \tmmathbf{\kappa}_i$ and $\delta^2 \tmmathbf{\kappa}_i$, are available
from Section~\ref{sec:variations}: the equilibrium can be solved using
numerical methods that require evaluations of the Hessian of the energy. Note
that the Hessian can be represented as a sparse matrix thanks to the local
nature of the energy contributions $E_i (\tmmathbf{\kappa}_i (\tmmathbf{x}_{i
- 1}, \varphi^{i - 1}, \tmmathbf{x}_i, \varphi^i, \tmmathbf{x}_{i + 1}))$ in
equation~(\ref{eq:discrete-strain-energy-density}).

In the applications presented in the forthcoming sections, we
find equilibrium configurations by minimizing
$\Phi (\tmmathbf{X})$ in equation~(\ref{eq:discrete-strain-energy-density})
using the sequential quadratic programming method (SQP) described by
{\cite{Nocedal2006}}; it is an extension of the Newton method for non-linear
optimization problems which can handle the non-linear constraints in
equation~(\ref{eq:elastic-simulations-inextensibility-cstraint}). It requires
the evaluation of the first and second gradient of the energy $\Phi$, see
equation~(\ref{eq:strain-energy-gradient-hessian}), and of the first gradient
of the constraints that are available from
equation~(\ref{eq:variations-epsiloni}). We used an in-house implementation of
the SQP method in the \tmverbatim{C++} language, with matrix inversion done
using the \tmverbatim{SimplicialLDLT} method available from the Eigen
library~{\cite{eigenweb}}.

\subsection{Inextensible elastic ribbons (Wunderlich
model)}\label{ssec:ribbons}

Ribbons made up of material that are sensitive to
light~{\cite{Yu-Nakano-EtAl-Photomechanics:-Directed-bending-2003,Gelebart-Mulder-EtAl-Making-waves-in-a-photoactive-2017}}
or temperature
change~{\cite{Bae-Na-EtAl-Edge-defined-metric-buckling-2014}} have
been used to design lightweight structures that can be actuated.  They
are easy to fabricate, typically by cutting out a thin sheet of
material, and their thin geometry can turn the small strains produced
by actuation into large-amplitude motion.  For this reason, there has
been a surge of interest towards mechanical models for elastic ribbons
recently.  When the width-to-thickness ratio of a ribbon cross-section
is sufficiently large, its mid-surface is effectively inextensible.
Sadowsky has proposed a one-dimensional mechanical model for
inextensible
ribbons~{\cite{Sadowsky-Die-Differentialgleichungen-des-Mobiusschen-Bandes-1929}}.
Sadowsky model is one-dimensional but differs from classical rod
models in two aspects: one of the two bending modes is inhibited due
to the large width-to-thickness aspect-ratio, and the two remaining
twisting and bending modes are governed by an {\tmem{non-quadratic}}
strain energy potential that effectively captures the inextensible
deformations of the ribbon mid-surface.  Sadowsky's strain energy is
non-convex which can lead to the formation of non-smooth solution
representing a
micro-structure~{\cite{Freddi-Hornung-EtAl-A-corrected-Sadowsky-functional-2015,Paroni-Tomassetti-Macroscopic-and-Microscopic-Behavior-2019}};
to avoid these difficulties, we use the higher-order model of
Wunderlich that accounts for the dependence of the energy on the
longitudinal gradient of bending and twisting
strain~{\cite{Wunderlich-Uber-ein-abwickelbares-Mobiusband-1962}}.

The Wunderlich model has been solved numerically by a continuation method, see
for instance the work
of~{\cite{Starostin-Heijden-Equilibrium-Shapes-with-2015}}. The continuation
method is an extension of the shooting method that can efficiently track
solutions depending on a
parameter~{\cite{Doedel-Champneys-EtAl-AUTO-07p:-continuation-and-bifurcation-2007}}.
It requires the full boundary-value problem of equilibrium to be specified
spelled out, which is quite impractical in the case of Wunderlich ribbons. A
recent and promising alternative is the high-order method
of~{\cite{Charrondiere-Bertails-Descoubes-EtAl-Numerical-modeling-of-inextensible-2020}}
that starts from linear and quadratic interpolations of the bending and
twisting strains, and treats the center-line position and the directors as
secondary (reconstructed) quantities. In the present work, we explore an
alternative approach, and show that simulations of the Wunderlich model are
possible with limited additional work on top of the generic Discrete elastic
rod framework.

We build  on the work
of~{\cite{Dias-Audoly-Wunderlich-meet-Kirchhoff:-2015}} who have shown that
the Wunderlich model can be viewed as a special type of a non-linear elastic
rod, see also~{\cite{Starostin-Heijden-Force-and-moment-balance-2009}}.
Accordingly, simulations of the Wunderlich model can be achieved using a
simple extension of the Discrete elastic rod model, which we describe now. We
first introduce a geometric model for a {\tmem{discrete inextensible
ribbon}},%
\begin{figure}
  \centerline{\includegraphics[width=.99\textwidth]{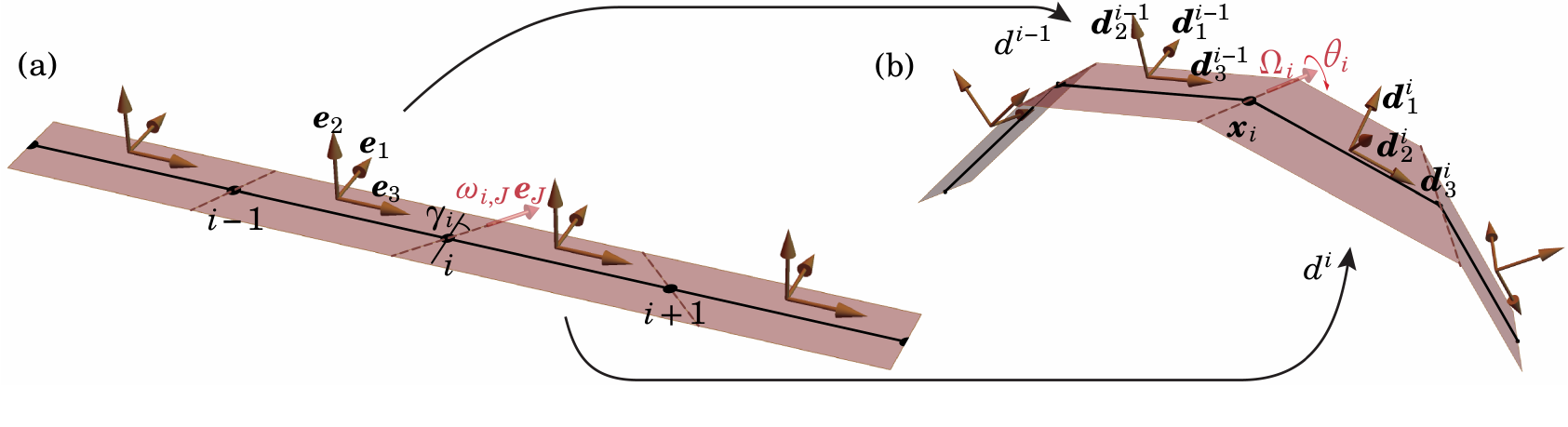}}
  \caption{A discrete inextensible ribbon: (a)~flat configuration and
  (b)~current (folded) configuration obtained by folding along the
  generatrices (brown dashed lines) by an angle $\theta_i$. By the
  inextensibility condition, the fold line through vertex $\tmmathbf{x}_i$ in
  current configuration lies at the intersection of the adjacent faces,
  {\tmem{i.e.}}, of the planes spanned by $\tmmathbf{d}_1^{i - 1}$ and
  $\tmmathbf{d}_3^{i - 1}$ on the one hand and by $\tmmathbf{d}_1^i$ and
  $\tmmathbf{d}_3^i$ on the other hand.\label{fig:inext-ribbon}}
\end{figure}
in which the inextensibility of the mid-surface is fully taken
into account. We start from a rectangular strip lying in the plane spanned by
$(\tmmathbf{e}_1, \tmmathbf{e}_3)$, as shown in
Figure~\ref{fig:inext-ribbon}a. Through every node (shown as black dots in the
figure), we pick a folding direction within the plane of the strip (brown
dotted line in the figure); we denote by $\mathpi / 2 - \gamma_i$ the angle of
the fold line relative to the centerline. Next, we fold along each one of
these lines by an angle $\theta_i$, as shown in
Figure~\ref{fig:inext-ribbon}b. We call  the
resulting surface a discrete inextensible ribbon. By construction, it is isometric to the original strip.

Let us now introduce the director frames $\tmmathbf{d}_I^i$ following rigidly
each one of the faces: the planar faces are spanned by the directors
$\tmmathbf{d}_1^i$ and $\tmmathbf{d}_3^i$. By construction the vector
$\tmmathbf{\Omega}_i$ for the rotation that maps one frame, $\tmmathbf{d}_I^{i
- 1}$, to the next, $\tmmathbf{d}_I^i$, see
equation~(\ref{eq:from-left-to-right-frame}), is aligned with the fold line.
We observe that the unit tangent along the fold direction is $\tmmathbf{e}_3
\tmmult \sin \gamma_i +\tmmathbf{e}_1 \tmmult \cos \gamma_i$ in the flat
configuration of the strip; it is therefore mapped to $\tmmathbf{d}_3^{i - 1}
\tmmult \sin \gamma_i +\tmmathbf{d}_1^{i - 1} \tmmult \cos \gamma_i
=\tmmathbf{d}_3^i \tmmult \sin \gamma_i +\tmmathbf{d}_1^i \tmmult \cos
\gamma_i$ in the current configuration. In view of this, we conclude
\[ \tmmathbf{\Omega}_i = \left( \tmmathbf{d}_3^{i - 1} \tmmult \sin \gamma_i
   +\tmmathbf{d}_1^{i - 1} \tmmult \cos \gamma_i \right) \tmmult \theta_i =
   \left( \tmmathbf{d}_3^i \tmmult \sin \gamma_i +\tmmathbf{d}_1^i \tmmult
   \cos \gamma_i \right) \tmmult \theta_i . \]
Comparing with equation~(\ref{eq:rotation-vector-from-left-to-right-frame}),
we obtain the discrete deformation measure in the developable ribbon as $\omega_{i, 1} =
\theta_i \tmmult \cos \gamma_i$ (bending mode), $\omega_{i, 2} = 0$ (inhibited
bending mode) and $\omega_{3, i} = 0$ (twisting mode). Eliminating $\theta_i$,
we find $\omega_{i, 2} = 0 \text{ and } \frac{\omega_{i, 3}}{\omega_{i, 1}} =
\tan \gamma_i$, which can be rewritten in terms of the original discrete
strain $\tmmathbf{\kappa}_i = (\kappa_{i, 1}, \kappa_{i, 2}, \kappa_{i, 3})$
with the help of equation~(\ref{eq:psi}) as
\begin{equation}
  \begin{array}{rll}
    \kappa_{i, 2} & = & 0\\
    \kappa_{i, 3} & = & \eta_i \tmmult \kappa_{i, 1}
  \end{array} \label{eq:discrete-developability-condition}
\end{equation}
where
\[ \eta_i = \tan \gamma_i . \]
The continuous version of the developability conditions is $\kappa_2 (s) = 0$
and $\kappa_3 (s) = \eta (s) \tmmult \kappa_1 (s)$, where $\eta (s) = \tan
\gamma (s)$ and $\mathpi / 2 - \gamma (s)$ is the angle between the generatrix
and the tangent, see for
instance~{\cite{Dias-Audoly-Wunderlich-meet-Kirchhoff:-2015}}. It is
remarkable that the discrete developability
conditions~(\ref{eq:discrete-developability-condition}) are 
identically satisfied. This is a consequence of the simple geometric interpretation for
the discrete deformation measures introduced in Section~\ref{sec:discrete-kinematics}.

To simulate inextensible ribbons, we introduce the unknown $\eta_i$ as an
additional degree of freedom at each one of the interior nodes, and we use in
equation~(\ref{eq:discrete-strain-energy-density}) a strain energy density
directly inspired from that of
Wunderlich~{\cite{Dias-Audoly-Wunderlich-meet-Kirchhoff:-2015,Starostin-Heijden-Equilibrium-Shapes-with-2015}}
\begin{equation}
  E_i (\tmmathbf{\kappa}_i, \eta_{i - 1}, \eta_i, \eta_{i + 1}) = \frac{D
  \tmmult w}{2 \tmmult \ell} \kappa_{i, 1}^2 \tmmult (1 + \eta_i^2)^2 \tmmult
  \frac{1}{w \tmmult \eta_i'} \tmmult \ln \left( \frac{1 + \frac{1}{2} \tmmult
  \eta_i' \tmmult w}{1 - \frac{1}{2} \tmmult \eta_i' \tmmult w} \right) .
  \label{eq:discrete-Wunderlich-energy}
\end{equation}
In equation~(\ref{eq:discrete-Wunderlich-energy}), $D = \frac{Y \tmmult
h^3}{12 \tmmult (1 - \nu^2)}$ is the bending modulus from plate theory, $h$ is
the thickness, $w$ is the width and $\ell$ is the discretization length. The
quantity $\eta_i'$ is calculated by a central-difference approximation of the
gradient of $\eta$,
\[ \eta_i' = \frac{\eta_{i + 1} - \eta_{i - 1}}{2 \tmmult \ell}, \]
where $\ell$ is the mesh size.
The constraint~(\ref{eq:discrete-developability-condition})$_{2}$ is
imposed at each node using the SQP method.  Introducing the nodal
degrees of freedom $\eta_i$ together with the constraint
(\ref{eq:discrete-developability-condition})$_{2}$ allows us to work
around calculating $\eta_i = \kappa_{i, 3} / \kappa_{i, 1}$, which is a
division with a potentially small denominator; in addition, this
approach warrants that $\kappa_{i, 3}=0$ whenever $\kappa_{i, 1}=0$,
which is necessary for the Wunderlich energy to remain finite.

It is a feature of the Wunderlich model that $\eta$ can take on
arbitrary values in intervals where $\kappa_1$ vanishes identically.
To work around this, we have introduced an artificial drag on the
$\eta_i$'s between iterations of the solve.  When convergence is reached, the drag force is identically zero.

The discrete potential energy $\Phi (\tmmathbf{X})$ is minimized by the same
numerical method as described in Section~\ref{sec:ribbons}\ref{ssec:linearly-elastic-rods}, taking
into account the kinematic
constraints~(\ref{eq:discrete-developability-condition}) and the centerline
inextensibility~(\ref{eq:elastic-simulations-inextensibility-cstraint}).

\section{Illustrations}
\label{sec:illustrations}

In this section, the Discrete elastic rod model is used to simulate
\begin{itemize}
  \item a linearly elastic model for an isotropic beam, 
  Section~\ref{sec:illustrations}\ref{ssec:illustration-Euler-buckling},
  
  \item a linearly elastic model for an anisotropic beam with natural
  curvature, Section~\ref{sec:illustrations}\ref{ssec:illustration-over-curvature},
  
  \item Sano and Wada's extensible ribbon model,
 Section~\ref{sec:illustrations}\ref{ssec:illustration-Sano-ribbon},
  
  \item Wunderlich's inextensible ribbon model,
  Section~\ref{sec:illustrations}\ref{ssec:illustration-Moebius-ribbon}.
\end{itemize}
These examples serve to illustrate the capabilities of the model. In addition,
comparison with reference solutions available from the literature provide a
verification of its predictions.

\subsection{Euler buckling}\label{ssec:illustration-Euler-buckling}

\begin{figure}
  \centerline{\includegraphics[width=.99\textwidth]{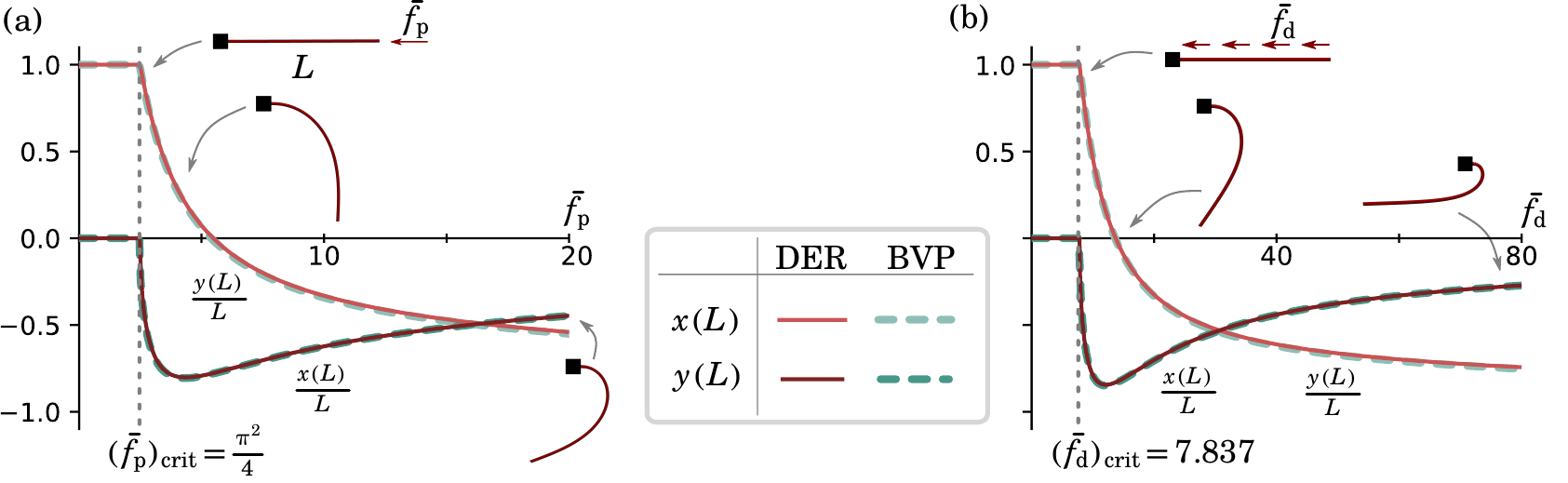}}
  \caption{Buckling of a planar Elastica subject to (a)~a point-like force
  applied at the endpoint and (b)~a distributed force. Comparison of the
  solutions of the boundary-value problem~(\ref{eq:Elastica-bvp}) by a
  numerical shooting method (dashed curves) and of the Discrete elastic rod
  method (solid curves): the scaled coordinates of the endpoint $s = L$ are
  plotted as a function of the dimensionless load. The dotted vertical line is
  the first critical load predicted by a linear bifurcation analysis from
  equation~(\ref{eq:Euler-linear-bifurcation}).\label{fig:Euler-buckling}}
\end{figure}

We consider Euler buckling for a planar, inextensible elastic rod that is
clamped at one endpoint. We consider two types of loading: either a point-like
force $f_{\text{p}}$ at the endpoint opposite to the clamp, or a force
$f_{\text{d}}$ distributed along the length of the rod. In both cases, the
force is applied along the initial axis of the rod, is invariable (dead
loading), and is counted positive when compressive. A sketch is provided in
Figure~\ref{fig:Euler-buckling}.

Mathematically, the equilibria of the rod having bending modulus $B$ are the
stationary points of the functional $\Phi = \int_0^L \frac{B}{2} \tmmult
{\theta'}^2 (s) \tmmult \mathd s + f_{\text{p}} \tmmult x (L)$ (point load) 
or $\Phi = \int_0^L \left( \frac{B}{2} \tmmult {\theta'}^2 (s) +
f_{\text{d}} \tmmult x (s) \right) \tmmult \mathd s$ (distributed load),
subject to the clamping condition $\theta (0) = 0$. The coordinates of a
point on the centerline $(x (s), y (s))$ are reconstructed using the
inextensibility condition as $x (s) \tmmult \tmmathbf{e}_1 + y (s) \tmmult
\tmmathbf{e}_2 = \int_0^s \left( \cos \theta \tmmult \tmmathbf{e}_1 + \sin
\theta \tmmult \tmmathbf{e}_2 \right) \tmmult \mathd s$.

The boundary-value equilibrium problem for the Elastica is obtained by the
Euler-Lagrange method as
\begin{equation}
  0 = B \tmmult \theta'' (s) + \sin \theta (s) \times \left\{\begin{array}{ll}
    f_{\text{p}} & \text{(point-like load)}\\
    f_{\text{d}} \tmmult (L - s) & \text{(distributed load)}
  \end{array}\right. \qquad \theta (0) = 0 \qquad \theta' (L) = 0.
  \label{eq:Elastica-bvp}
\end{equation}
By writing this problem in dimensionless form, one can effectively set the
bending modulus, the length and the load to $B = 1$, $L = 1$, and
$f_{\text{p}} = \overline{f}_{\text{p}}$ (point-like load) or $f_{\text{d}} =
\overline{f}_{\text{d}}$ (distributed load), where the dimensionless load is
\begin{equation}
  \overline{f}_{\text{p}} = \frac{f_{\text{p}}}{B / L^2} \qquad
  \overline{f}_{\text{d}} = \frac{L \tmmult f_{\text{d}}}{B / L^2} .
  \label{eq:Euler-scaled-load}
\end{equation}
The critical buckling loads are found by solving the {\tmem{linearized}}
version of the buckling problem~(\ref{eq:Elastica-bvp}) (linear bifurcation
analysis),
\begin{equation}
  \begin{array}{rlll}
    \left( \overline{f}_{\text{p}} \right)_{\text{crit}} & = &
    \frac{\mathpi^2}{4} & \text{\quad (point-like load)}\\
    \left( \overline{f}_{\text{d}} \right)_{\text{crit}} & = & 7.837 &
    \text{\quad (distributed load)}
  \end{array} \label{eq:Euler-linear-bifurcation}
\end{equation}

Numerical simulations of this Euler buckling problem are conducted
using the Discrete elastic rod method, as explained in
Section~\ref{eq:discrete-strain-energy-density}.  Simulations are set
up with $B = 1$, $L = 1$, number of nodes $N = 100$.  In view of this
we expect to the buckling loads to be $f_{\text{d}} =
\overline{f}_{\text{d}}$ $f_{\text{p}} = \overline{f}_{\text{p}}$.
The inextensibility constraint is enforced exactly using SQP. 
The clamped boundary is enforced by fixing the first and second
nodes as well as the first frame. 

The typical simulation time is about 1/10s for each equilibrium on a
personal computer, and the results are shown in
Figure~\ref{fig:Euler-buckling}, and compared to that obtained by
solving~(\ref{eq:Elastica-bvp}) using the \tmverbatim{bvp4c} solver
from Matlab.  A good agreement on the position of the endpoint of the
rod is found in the entire post-bifurcation regime.  In addition, the
onset of bifurcation agrees accurately with the
prediction~(\ref{eq:Euler-linear-bifurcation}) from the linear
stability analysis.

\subsection{Folding of an over-curved
ring}\label{ssec:illustration-over-curvature}

A circular elastic ring with length $L$ can buckle out of plane if its natural
natural curvature $\kappa_{(0)}$ does not match the curvature $2 \mathpi / L$
of the circle with length $L$. In the case of an over-curved ring, such that
$\kappa_{(0)} > 2 \tmmult \mathpi / L$, a buckled shape featuring two
symmetric lobes has been
reported~{\cite{Mouthuy-Coulombier-EtAl-Overcurvature-describes-the-buckling-2012,Bae-Na-EtAl-Edge-defined-metric-buckling-2014,Audoly-Seffen-Buckling-of-naturally-curved-2015}}.
Here, we simulate the buckling of over-curved rings using the Discrete elastic
rod model and compare the results to the experimental shapes reported
by~{\cite{Mouthuy-Coulombier-EtAl-Overcurvature-describes-the-buckling-2012}}.

In the experiments
of~{\cite{Mouthuy-Coulombier-EtAl-Overcurvature-describes-the-buckling-2012}},
a commercial slinky spring with a width $w = 5 \text{~$\tmop{mm}$}$, thickness
$t = \text{2~$\tmop{mm}$}$ and length $L = 314 \text{~mm}$ is used; Poisson's
ratio has been measured as $\nu = 0.41$. Note that the aspect-ratio $t / w =
0.4$ is not small. In our simulations, we use a discrete version of the
quadratic strain energy for a linearly elastic rod having an anisotropic
cross-section ($I_1 \neq I_2$), see
equations~(\ref{eq:strain-energy-linearly-elastic}--\ref{eq:discrete-strain-energy-density}).
We use the elastic moduli reported in the supplement
of~{\cite{Mouthuy-Coulombier-EtAl-Overcurvature-describes-the-buckling-2012}}:
\begin{equation}
  Y \tmmult I_1 = Y \tmmult \frac{w \tmmult t^3}{12} \qquad Y \tmmult I_2 = Y
  \tmmult \frac{w^3 \tmmult t}{12} \qquad \mu \tmmult J = Y \tmmult
  \frac{0.256 \tmmult w \tmmult t^3}{2 \tmmult (1 + \nu)} .
  \label{eq:mouthy-linear-moduli}
\end{equation}
The value $0.256$ in the numerator was obtained
by~{\cite{Mouthuy-Coulombier-EtAl-Overcurvature-describes-the-buckling-2012}}
from the book
of~{\cite{Ugural-Fenster-Advanced-mechanics-of-materials-2019}}, and applies
to the particular commercial Slinky used in their experiments. In the absence
of applied loading, the value of the Young modulus is irrelevant and we set $Y
= 1$ in the simulations.

\begin{figure}
  \centerline{\includegraphics[width=.75\textwidth]{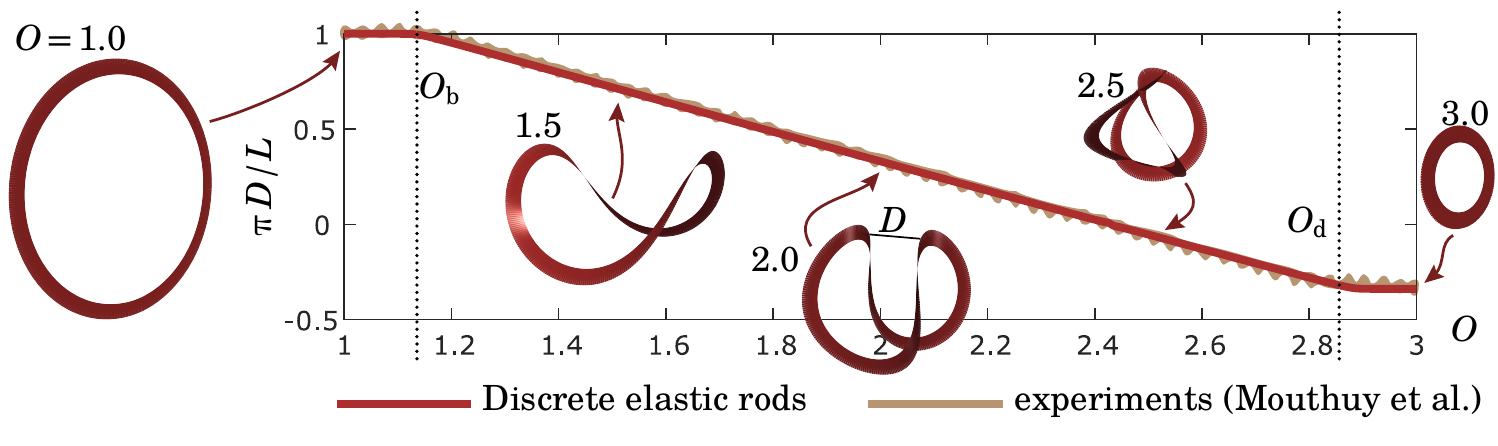}}
  \caption{Equilibrium of an over-curved elastic ring. Material and geometric
  parameters correspond to the slinky used
  by~{\cite{Mouthuy-Coulombier-EtAl-Overcurvature-describes-the-buckling-2012}}
  (see main text for values). a)~Equilibrium configurations for different
  values of the over-curvature ratio $O$. b)~Minimal distance of
  approach $D$ as a function of $O$: comparison of Discrete elastic rod
  simulations and
  experiments~{\cite{Mouthuy-Coulombier-EtAl-Overcurvature-describes-the-buckling-2012}}.
  The simulations reproduces both the initial buckling at $O_{\text{b}}$, and
  the `de-buckling' into a flat, triply covered ring at
  $O_{\text{d}}$.\label{fig:overcurvedconfig}}
\end{figure}

The equilibria of the Discrete elastic rod are calculated numerically for
different values of the dimensionless loading parameter $O = 2 \tmmult \pi
\tmmult \kappa_{(0)} / L$, with $O > 1$ corresponding to the over-curved case.
We use $N = 400$ nodes. We start from a circular configuration having
curvature $\kappa_{(0)} = 2 \tmmult \mathpi / L$. The Discrete elastic rod
model is closed into a ring as follows: the first two nodes and the last two
nodes are prescribed to $\tmmathbf{x}_0 =\tmmathbf{x}_{N - 1} =\tmmathbf{0}$
and $\tmmathbf{x}_1 =\tmmathbf{x}_N = \ell \tmmult \tmmathbf{e}_x$,
respectively; the first and last frames are also fixed, such that
$\tmmathbf{d}_1^0 =\tmmathbf{d}_1^{N - 1} =\tmmathbf{e}_y$. Next, the
over-curvature $\kappa_{(0)}$ is varied incrementally. For each value of
$\kappa_{(0)}$, an equilibrium configuration is sought, and we extract the
minimal distance $D$ between pairs of opposite points on the ring. In
Figure~\ref{fig:overcurvedconfig}, the scaled distance $D$ is plotted as a
function of $O$. A good agreement is found with the experiments over the
entire range of values of the over-curvature parameter $O > 1$. The
simulations correctly predict a planar, triply covered circular solution for
$O > O_{\text{d}} \approx 2.85$, as seen in the experiments.

\subsection{Buckling of a bent and twisted
ribbon}\label{ssec:illustration-Sano-ribbon}

We now turn to an effective rod model applicable to thin ribbons.
Sano and Wada {\cite{Sano-Wada-Twist-Induced-Snapping-in-a-Bent-2019}}
have proposed an effective beam model that accounts for the
stretchability of the ribbon having moderate width, thereby improving
on Sadowsky's inextensibility assumption.  A discrete version of their
continuous model is of the
form~(\ref{eq:discrete-strain-energy-density}) with a strain energy
per elastic hinge
\begin{equation}
  E_i (\kappa_1, \kappa_2, \kappa_3) = \frac{1}{2 \tmmult \ell} \tmmult \left(
  A_1 \tmmult \kappa_1^2 + A_2 \tmmult \left( \kappa_2^2 +
  \frac{\kappa_3^4}{\ell^2 / \xi^2 + \kappa_2^2} \right) + A_3 \tmmult
  \kappa_3^2 \right) . \label{eq:Ei-Sano}
\end{equation}
Here, $\ell$ is the uniform segment length in undeformed configuration, $A_1 =
Y \tmmult h \tmmult w^3 / 12$ and $A_2 = Y \tmmult h^3 \tmmult w / 12$ are the
initial bending moduli, $A_3 = Y \tmmult h^3 \tmmult w / [6 \tmmult (1 +
\nu)]$ is the initial twisting modulus and $\xi^2 = (1 - \nu^2) \tmmult w^4 /
60 \tmmult h^2$. The parameter $\xi$ is the typical length-scale where the
stretchability of the mid-surface starts to play a role. The potential $E_i$
from equation~(\ref{eq:Ei-Sano}) is non-quadratic, meaning that the equivalent
rod has non-linear elastic constitutive laws.

The elastic model~(\ref{eq:Ei-Sano}) of Sano and Wada is applicable to
thin ribbons, for $w \gg h$.  It is based on kinematic approximations.
A refined version of their model has been obtained very recently
by~{\cite{Audoly-Neukirch-A-one-dimensional-model-for-elastic-2021}},
by asymptotic expansion starting from shell theory; in the latter
work, a detailed discussion of the validity of the various models for
thin ribbons can also be found.  We do not expect any difficulty in
applying the present numerical model to the ribbon model
in~{\cite{Audoly-Neukirch-A-one-dimensional-model-for-elastic-2021}}.
Both the models of Sano and Wada, and of Audoly and Neukirch improve
on Wunderlich model by addressing the stretchability of the ribbon;
unlike the Wunderlich model, however, they ignore the dependence of
the energy on $\eta'$, and therefore account less accurately for the
`conical' singularities often observed in
ribbons~\cite{Yu-Hanna-Bifurcations-of-buckled-clamped-2019} as
$\eta$ varies quickly there.

\begin{figure}
  \centerline{\includegraphics[width=.99\textwidth]{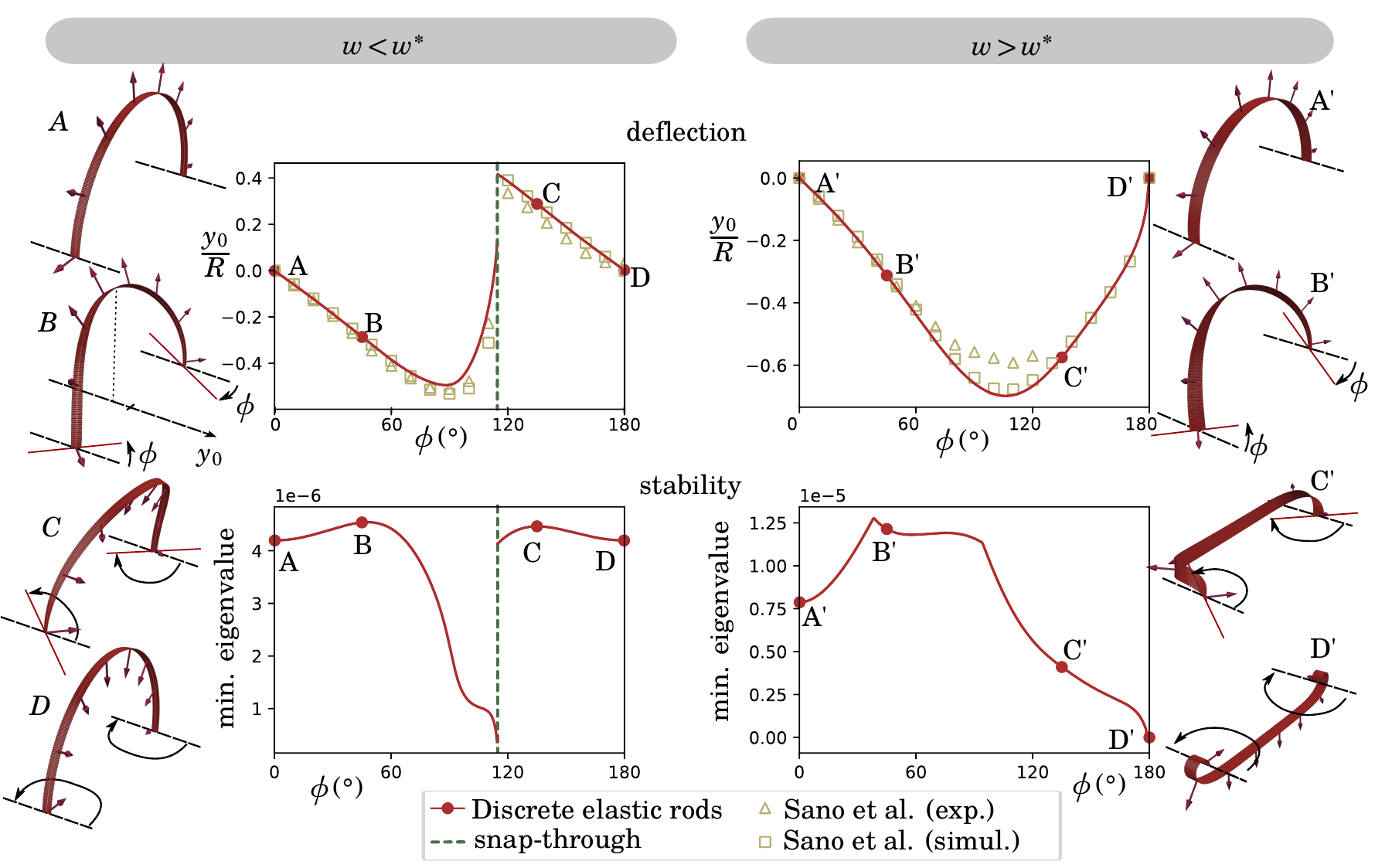}}
  \caption{Equilibria of an extensible ribbon, as captured by Sano and Wada's
  equivalent rod model, see equation~(\ref{eq:Ei-Sano}). {\tmem{Top row}}:
  equilibrium diagram showing the scaled value of the deflection $y_0$ at the
  center of the ribbon as a function of the twisting angle $\phi$ at the
  endpoints. Comparison of the experiments (triangles) and simulations
  (squares) from {\cite{Sano-Wada-Twist-Induced-Snapping-in-a-Bent-2019}}
  with simulations using the Discrete elastic rod model (solid curves and
  circles). {\tmem{Left column}}:~moderately wide ribbon $(h, w, R) = (0.2, 8,
  108)$\text{mm} showing a snapping instability; {\tmem{Right column}}: wider
  ribbon $(h, w, R) = (0.2, 15, 108)$\text{mm}, in which the instability is
  suppressed. {\tmem{Bottom row}}: smallest eigenvalues of the tangent
  stiffness matrix, on the same solution branch shown as shown in the plot
  immediately above: the presence of an instability for $w < w^{\asterisk}$
  (left column) is confirmed by the fact that the smallest eigenvalue reaches
  zero when the instability sets in.\label{fig:sanocomp}}
\end{figure}

Following {\cite{Sano-Wada-Twist-Induced-Snapping-in-a-Bent-2019}}, we 
consider the buckling of a ribbon with length $L = \mathpi \tmmult
R$ bent into half a circle, whose ends are twisted in an opposite senses by an
angle $\phi$, see Figure~\ref{fig:sanocomp}. Specifically, they identified a
snapping instability which occurs for moderately wide ribbons, when the width
$w < w^{\asterisk}$ is below a threshold $w^{\asterisk} \approx 1.24 \tmmult
\sqrt{L \tmmult h}$, but not for wider ribbons, when $w > w^{\asterisk}$; they
showed that their equivalent rod model can reproduce this instability, as well
as its disappearance for larger widths. In Figure~\ref{fig:sanocomp}, we
compare the predictions of a Discrete elastic rod model using~(\ref{eq:Ei-Sano})
with the original experiments and simulations
from~{\cite{Sano-Wada-Twist-Induced-Snapping-in-a-Bent-2019}}. Our
simulations use $N = 350$ vertices each. Our simulation results are in close
agreement with both their experimental and numerical results. In particular,
we recover the instability when $w < w^{\asterisk}$
only.

\subsection{The elastic M{\"o}bius
band}\label{ssec:illustration-Moebius-ribbon}

An extension of the Discrete elastic rod model that simulates the inextensible
ribbon model of Wunderlich has been described in Section~\ref{sec:ribbons}\ref{ssec:ribbons},
see equation~(\ref{eq:discrete-Wunderlich-energy}). With the aim
to illustrate and verify this discrete model, we simulate the equilibrium of a
M{\"o}bius ribbon, and compare the results to those reported in the seminal
paper of Starostin and van der Heijden {\cite{Starostin-Heijden-The-shape-of-a-Mobius-strip-2007}}. In our
simulations, the inextensible strip is first bent into a circle, and the
endpoints are turned progressively twisted by an angle of $180^{\circ}$ to provide the correct topology. The final equilibrium
shapes are then recorded for all possible values of the aspect-ratio $w / L$.
For these final equilibrium shapes, the conditions $\tmmathbf{x}_0
=\tmmathbf{x}_{N - 1} =\tmmathbf{0}$ and $\tmmathbf{x}_1 =\tmmathbf{x}_N =
\ell \tmmult \tmmathbf{e}_x$ hold as earlier, and the orientation of the
terminal material frames are such that \ $\tmmathbf{d}_1^0 = +\tmmathbf{e}_y$
and $\tmmathbf{d}_1^{N - 1} = -\tmmathbf{e}_y$.

\begin{figure}
  \centerline{\includegraphics[width=.99\textwidth]{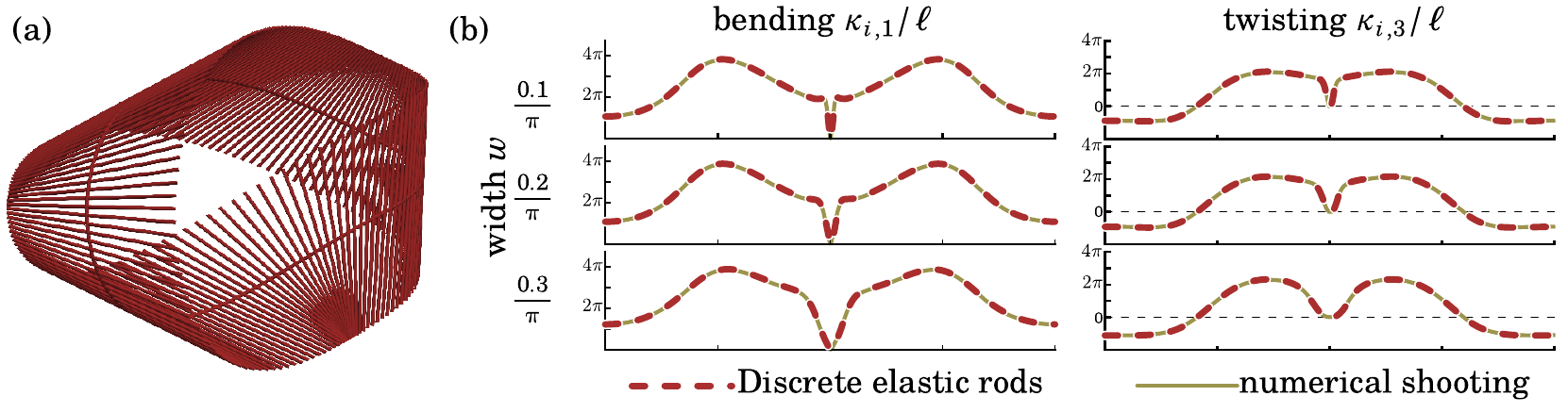}}
  \caption{Simulation of an inextensible M{\"o}bius strip with $L = 1$.
  (a)~Equilibrium width $w = 1 / \left( 2 \tmmult \mathpi \right)$, as
  simulated by the Discrete elastic rod model from Section~\ref{sec:ribbons}\ref{ssec:ribbons}
  with $N = 150$ nodes. (b)~Distribution of bending and twisting: Discrete
  elastic rod simulations with $N = 250$ vertices (dashed curves) versus
  solution of {\cite{Starostin-Heijden-The-shape-of-a-Mobius-strip-2007}}
  obtained by numerical shooting (solid curves); the latter have been properly
  rescaled to reflect our
  conventions.\label{fig:mobiusstrip}}
\end{figure}

The equilibrium shape for a particular aspect-ratio $w / L = 1 / \left( 2
\tmmult \mathpi \right)$ is shown in Figure~\ref{fig:mobiusstrip}a, with
arc-length $L = 1$, width $w = 1 / \left( 2 \tmmult \mathpi \right)$ and $N =
150$ simulation nodes. A detailed comparison with the results of
{\cite{Starostin-Heijden-The-shape-of-a-Mobius-strip-2007}} is provided in
Figure~\ref{fig:mobiusstrip}b, where the scaled bending and twisting strains
$\kappa_{i, 1} / \ell$ and $\kappa_{i, 3} / \ell$ from the discrete model with
$N = 250$ vertices are compared to the strains $\kappa_1 (s)$ and $\kappa_3
(s)$ obtained by {\cite{Starostin-Heijden-The-shape-of-a-Mobius-strip-2007}}
using numerical shooting, for different values of the width $w$.

\section{Conclusion}

We have presented a new formulation of the Discrete elastic rod model.
The formulation is concise and uses only the minimally necessary
degrees of freedom: the position of the nodes and the angle of twist
of the segments between the nodes.  It naturally incorporates the
adaptation condition without the need for any constraint, penalty or
Lagrange multiplier.  We use bending and twisting deformation measures
that are different from those used in earlier work on Discrete elastic
rods, are equally consistent with their continuum counterparts, and
have a simple physical interpretation in the discrete setting.
Consequently, the formulation is versatile in the sense that it can be
combined with a variety of linear and nonlinear as well as elastic and
inelastic constitutive relations.  In fact, ribbons can be
incorporated as generalized rods with a nonlinear constitutive model.
Similarly, the formulation can be used both for static and dynamic
simulations.

We have  presented explicit formulae for the first and second derivatives of the deformation
measures that eases implementation.  We have demonstrated our method with four
examples, and verified our results against prior experimental and theoretical findings in the literature.

\enlargethispage{20pt}

The source code used for the numerical simulation is available through CaltechDATA at \url{https://data.caltech.edu/records/2147}.

All three authors conceived of the work and the formulation.  KK conducted the theoretical and numerical calculations with advice from BA and KB.  KK and BA took the lead in writing the manuscript and all three authors finalized it.

The authors declare that there are no competing interests.

The work began when BA visited Caltech as a Moore Distinguished Scholar.  KK and KB gratefully acknowledge the support of the US Office of Naval Research through Multi-investigator University Research Initiative
Grant ONR N00014-18-1-2624.

\newpage
\appendix

\centerline{\Large{\bf Appendices}}

\section{Plot of function $\psi$}
\begin{figure}[b]
  \centerline{\includegraphics{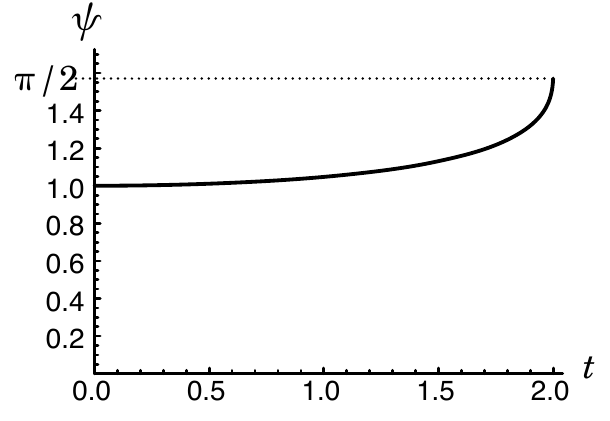}}
  \caption{Function $\psi (t)$ from equation~(\ref{eq:psi}) used to adjust the
  norm of the strain $\tmmathbf{\kappa}_i$ with $t = | \tmmathbf{\kappa}_i |$,
  see equation~(\ref{eq:from-kappai-to-omegai}).\label{fig:psi}}
\end{figure}

The function $\psi(t)$ from equation~(\ref{eq:psi}) is plotted in 
figure~\ref{fig:psi}.

\section{Detailed derivation of the strain
gradients}\label{app:variations}

In this appendix, we provide a detailed derivation of the first and second
gradients of the strain appearing in section~\ref{sec:variations}.

To derive the first gradient, we continue to use the conventions of
section~\ref{sec:variations}: we use a perturbation $\delta \tmmathbf{X}$ of
the degrees of freedom, and we denote by $\delta \tmmathbf{y}=\tmmathbf{f}'
(\tmmathbf{x}) \cdot \delta \tmmathbf{x}$ the first variation of a generic
quantity $\tmmathbf{y}=\tmmathbf{f} (\tmmathbf{x})$ entering in the
reconstruction of the discrete strain, where $\tmmathbf{x}$ depends indirectly
on the degrees of freedom $\tmmathbf{X}$.

For the second variation, however, we work here in a slightly more general
setting than in the main text, as we consider two {\tmem{independent}}
perturbations $\delta_1 \tmmathbf{X}$ and $\delta_2 \tmmathbf{X}$ of the
degrees of freedom. We denote by $\delta_1 \tmmathbf{x}$ and $\delta_2
\tmmathbf{x}$ the corresponding perturbations to the variable $\tmmathbf{x}$,
and by $\delta_1 \tmmathbf{y}$ and $\delta_2 \tmmathbf{y}$ the first-order
variations of the functions: $\delta_1 \tmmathbf{y}=\tmmathbf{f}'
(\tmmathbf{x}) \cdot \delta_1 \tmmathbf{x}$ and $\delta_2
\tmmathbf{y}=\tmmathbf{f}' (\tmmathbf{x}) \cdot \delta_2 \tmmathbf{x}$ are
simply obtained by replacing the generic increment $\delta \tmmathbf{x}$
appearing in the first order variation $\delta \tmmathbf{y}$ with $\delta_1
\tmmathbf{x}$ and $\delta_2 \tmmathbf{x}$, respectively. To obtain the second
variation, we perturb the argument $\tmmathbf{x}$ appearing in $\delta_1
\tmmathbf{y}=\tmmathbf{f}' (\tmmathbf{x}) \cdot \delta_1 \tmmathbf{x}$ as
$\tmmathbf{x}+ \delta_2 \tmmathbf{x}$, leaving $\delta_1 \tmmathbf{x}$
untouched, and we expand the result to first order in $\delta_2 \tmmathbf{x}$.
This yields a quantity denoted as $\delta_{12} \tmmathbf{y}$, which we can
write formally as $\delta_{12} \tmmathbf{y}=\tmmathbf{f}'' (\tmmathbf{x}) \of
(\delta_1 \tmmathbf{x} \otimes \delta_2 \tmmathbf{x})$, where $\tmmathbf{f}''
(\tmmathbf{x})$ is the Hessian. By a classical result in the calculus of
variations, the quantity $\delta_{12} \tmmathbf{y}$ is bilinear and symmetric
with respect to $\delta_1 \tmmathbf{x}$ and $\delta_2 \tmmathbf{x}$. The
second variation $\delta^2 \tmmathbf{y}$ given in the main text is the
quadratic form obtained by ultimately condensing the variations $\delta_1
\tmmathbf{x}$ and $\delta_2 \tmmathbf{x}$ appearing in $\delta_{12}
\tmmathbf{y}$ into a single perturbation $\delta \tmmathbf{x}= \delta_1
\tmmathbf{x}= \delta_2 \tmmathbf{x}$.

\subsection{Infinitesimal rotation vectors}

As an important preliminary result, we show that the first variation of a
rotation represented by a unit quaternion $s$ can be characterized by means of
first-order {\tmem{vector-valued}} increment $\delta \hat{\tmmathbf{s}} \in
\mathbb{R}^3$, and that the second variation of $s$ can be represented by
means of a second-order {\tmem{vector-valued}} increment $\delta_{12}
\hat{\tmmathbf{s}} \in \mathbb{R}^3$. These vectors will be referred as the
{\tmem{infinitesimal rotation vectors}}. They are connected to the variations
$\delta s$ and $\delta_{12} s$ of the quaternion by
\begin{equation}
  \begin{array}{rll}
    \delta s  & = & \frac{1}{2} \tmmult \delta \hat{\tmmathbf{s}} 
	\tmmult s\\
    \delta_{12} s & = & \Bigl( \frac{1}{2} \tmmult \delta_{12}
    \hat{\tmmathbf{s}} - \frac{1}{4} \tmmult \delta_1 \hat{\tmmathbf{s}} \cdot
    \delta_2 \hat{\tmmathbf{s}} \Bigr) \tmmult s.
  \end{array} \label{eq:quaternion-increment-from-vector}
\end{equation}
The increment $\delta \hat{\tmmathbf{s}}$ is linear with respect to the
variation $\delta \tmmathbf{X}$ of the degrees of freedom, and the increment
$\delta_{12} \hat{\tmmathbf{s}}$ is bilinear with respect to the independent
variations $\delta_1 \tmmathbf{X}$ and $\delta_2 \tmmathbf{X}$ of the degrees
of freedom. As usual in our notation, $\delta_1 \hat{\tmmathbf{s}}$ and
$\delta_2 \hat{\tmmathbf{s}}$ denote the first-order variation $\delta
\hat{\tmmathbf{s}}$, evaluated on the increment $\delta_1 \tmmathbf{X}$ and
$\delta_2 \tmmathbf{X}$, respectively. This representation of the first and
second variations of a parameterized quaternion is equivalent to that proposed
by ~{\cite{Barbic-Zhao-Real-Time-Large-Deformation-Substructuring-2011}}.

The proof is as follows. By taking the first variation of the condition $2
\tmmult \left( s \tmmult \overline{s} - 1 \right) = 0$ that $s$ is a unit
quaternion, we have $0 = 2 \tmmult \delta s \tmmult \overline{s} + 2 \tmmult s
\tmmult \overline{\delta s} = 2 \tmmult \delta s \tmmult \overline{s} +
\overline{2 \tmmult \delta s \tmmult \overline{s}}$. This shows that the
quaternion $2 \tmmult \delta s \tmmult \overline{s}$ is a pure vector: this
the vector $\delta \hat{\tmmathbf{s}}$ introduced in
equation~(\ref{eq:quaternion-increment-from-vector}) above. Now, by inserting
the increment $\delta_1 \tmmathbf{X}$ in the relation just derived, we have $2
\tmmult \delta_1 s \tmmult \overline{s} \in \mathbb{R}^3$; perturbing this
expression as $s \leftarrow s + \delta_2 s$, one shows that the following
quaternion is a pure vector: $2 \tmmult \delta_{12} s \tmmult \overline{s} + 2
\tmmult \delta_1 s \tmmult \overline{\delta_2 s} = 2 \tmmult \delta_{12} s
\tmmult \overline{s} + \frac{1}{2} \tmmult \left( \delta_1 \hat{\tmmathbf{s}}
\tmmult s \right) \tmmult \overline{\left( \delta_2 \hat{\tmmathbf{s}} \tmmult
s \right)} = 2 \tmmult \delta_{12} s \tmmult \overline{s} - \frac{1}{2}
\tmmult \delta_1 \hat{\tmmathbf{s}} \tmmult \delta_2 \hat{\tmmathbf{s}} = 2
\tmmult \delta_{12} s \tmmult \overline{s} + \frac{1}{2} \tmmult \delta_1
\hat{\tmmathbf{s}} \cdot \delta_2 \hat{\tmmathbf{s}} - \frac{1}{2} \tmmult
\delta_1 \hat{\tmmathbf{s}} \times \delta_2 \hat{\tmmathbf{s}}$; here, the
quaternion product $\delta_1 \hat{\tmmathbf{s}} \tmmult \delta_2
\hat{\tmmathbf{s}}$ has been evaluated using the
definition~(\ref{eq:quaternion-mult}). Adding the vector quantity $\frac{1}{2}
\tmmult \delta_1 \hat{\tmmathbf{s}} \times \delta_2 \hat{\tmmathbf{s}}$, the
quantity $2 \tmmult \delta_{12} s \tmmult \overline{s} + \frac{1}{2} \tmmult
\delta_1 \hat{\tmmathbf{s}} \cdot \delta_2 \hat{\tmmathbf{s}}$ appears to be
another pure vector: this is the vector $\delta_{12} \hat{\tmmathbf{s}}$
introduced in equation~(\ref{eq:quaternion-increment-from-vector}).

The second-order infinitesimal rotation vector $\delta_{12}
\hat{\tmmathbf{s}}$ can be calculated directly from the first-order one
$\delta \hat{\tmmathbf{s}}$ as
\begin{equation}
  \delta_{12} \hat{\tmmathbf{s}} = \frac{\delta_1 (\delta_2
  \hat{\tmmathbf{s}}) + \delta_2 (\delta_1 \hat{\tmmathbf{s}})}{2} .
  \label{eq:2nd-order-infinitesimal-rotation-practical}
\end{equation}
Here, $\delta_1 (\delta_2 \hat{\tmmathbf{s}})$ denotes the first-order
variation of $\delta_2 \hat{\tmmathbf{s}}$ when $s$ is perturbed into $s +
\delta_1 s$; this quantity is {\tmem{not}} symmetric with respect to the
perturbations $\delta_1 s$ and $\delta_2 s$. Similarly, $\delta_2 (\delta_1
\hat{\tmmathbf{s}})$ denotes the first-order variation of $\delta_1
\hat{\tmmathbf{s}}$ when $s$ is perturbed into $s + \delta_2 s$.

The proof of equation~(\ref{eq:2nd-order-infinitesimal-rotation-practical}) is
as follows. Take the second variation of $\delta_1 s = \frac{1}{2} \tmmult
\delta_1 \hat{\tmmathbf{s}} \tmmult s$ from
equation~(\ref{eq:quaternion-increment-from-vector}) as $\delta_{12} s =
\frac{1}{2} \tmmult \delta_2 (\delta_1 \hat{\tmmathbf{s}}) \tmmult s +
\frac{1}{4} \tmmult \delta_1 \hat{\tmmathbf{s}} \tmmult \delta_2
\hat{\tmmathbf{s}} \tmmult s = \left( \frac{1}{2} \tmmult \delta_2 (\delta_1
\hat{\tmmathbf{s}}) - \frac{1}{4} \tmmult \delta_1 \hat{\tmmathbf{s}} \cdot
\delta_2 \hat{\tmmathbf{s}} + \frac{1}{4} \tmmult \delta_1 \hat{\tmmathbf{s}}
\times \delta_2 \hat{\tmmathbf{s}} \right) \tmmult s$. The left-hand side is
symmetric with respect to the perturbations $\delta_1 s$ and $\delta_2 s$, by
definition of the second variation. Symmetrizing the right-hand side, we
obtain $\delta_{12} s = \left( \frac{\delta_1 (\delta_2 \hat{\tmmathbf{s}}) +
\delta_2 (\delta_1 \hat{\tmmathbf{s}})}{4} - \frac{\delta_1 \hat{\tmmathbf{s}}
\cdot \delta_2 \hat{\tmmathbf{s}}}{4} \right) \tmmult s$. The infinitesimal
rotation vector $\delta_{12} \hat{\tmmathbf{s}}$ can then be identified from
equation~(\ref{eq:quaternion-increment-from-vector}), which yields the result
stated in equation~(\ref{eq:2nd-order-infinitesimal-rotation-practical}).

In the following sections, the first and second variations of the rotations
that enter into the Discrete elastic rod model, such as the parallel transport
$p^i$ and the director rotation $d^i$, will be systematically represented
using the corresponding infinitesimal rotation vectors, such as $\delta
\hat{\tmmathbf{p}}^i$, $\delta_{12} \hat{\tmmathbf{p}}^i$, $\delta
\hat{\tmmathbf{d}}^i$ and $\delta_{12} \hat{\tmmathbf{d}}^i$.

\subsection{Variation of parallel transport}

We start by deriving the variations of the parallel transport
$p_{\tmmathbf{a}}^{\tmmathbf{b}}$ from the unit vector $\tmmathbf{a}$ to the
unit vector $\tmmathbf{b}$ defined in equation~(\ref{eq: paralleltransport}),
assuming $\tmmathbf{b} \neq -\tmmathbf{a}$. As $\tmmathbf{a}$ represents the
fixed unit tangent $\tmmathbf{T}^i$ in reference configuration, it remains
unperturbed,
\[ \delta \tmmathbf{a}=\tmmathbf{0} \qquad \delta_{12}
   \tmmathbf{a}=\tmmathbf{0}. \]
Since $\tmmathbf{b}$ remains a unit vector during the perturbation, we have
$\frac{1}{2} \tmmult (| \tmmathbf{b} |^2 - 1) = 0$. Taking the first and
second variation of this constraint, we have
\[ \tmmathbf{b} \cdot \delta \tmmathbf{b}= 0 \qquad \tmmathbf{b} \cdot
   \delta_{12} \tmmathbf{b}+ \delta_1 \tmmathbf{b} \cdot \delta_2
   \tmmathbf{b}=\tmmathbf{0}. \]

\subsubsection{First variation of parallel transport}

As a preliminary step, we consider the case of parallel transport from
$\tmmathbf{b}$ to its perturbation $\tmmathbf{b}+ \delta \tmmathbf{b}$. Using
$\tmmathbf{b} \cdot \delta \tmmathbf{b}= 0$, we find from equation~(\ref{eq:
paralleltransport}),
\[ p_{\tmmathbf{b}}^{\tmmathbf{b}+ \delta \tmmathbf{b}} = 1 +
   \frac{\tmmathbf{b} \times \delta \tmmathbf{b}}{2} +\mathcal{O} (| \delta
   \tmmathbf{b} |^2) . \]

We now return to the calculation of $p_{\tmmathbf{a}}^{\tmmathbf{b}+ \delta
\tmmathbf{b}}$. Following the work of
{\cite{Bergou-Wardetzky-EtAl-Discrete-Elastic-Rods-2008}}, as well as
equations [3.7] and [A.2]
from~{\cite{Lestringant-Audoly-EtAl-A-discrete-geometrically-exact-2020}},
one can use a holonomy reasoning to shows that, to first order in $\delta
\tmmathbf{b}$,
\[ p_{\tmmathbf{a}}^{\tmmathbf{b}+ \delta \tmmathbf{b}} =
   p_{\tmmathbf{b}}^{\tmmathbf{b}+ \delta \tmmathbf{b}} \tmmult
   p_{\tmmathbf{a}}^{\tmmathbf{b}} \tmmult r_{\tmmathbf{a}} \left( -
   \frac{\tmmathbf{a} \times \tmmathbf{b}}{1 +\tmmathbf{a} \cdot \tmmathbf{b}}
   \cdot \delta \tmmathbf{b} \right) +\mathcal{O} (| \delta \tmmathbf{b} |^2)
   . \]
We rewrite this as
\begin{equation}
  p_{\tmmathbf{a}}^{\tmmathbf{b}+ \delta \tmmathbf{b}} =
  p_{\tmmathbf{b}}^{\tmmathbf{b}+ \delta \tmmathbf{b}} \tmmult
  p_{\tmmathbf{a}}^{\tmmathbf{b}} \tmmult r_{\tmmathbf{a}} (\delta \theta)
  +\mathcal{O} (| \delta \tmmathbf{b} |^2),
  \label{eq:variation-parallel-trsp-temp1}
\end{equation}
where $\delta \theta = - \frac{\tmmathbf{k}}{2} \cdot \delta \tmmathbf{b}$ and
$\tmmathbf{k}$ is the scaled binormal that characterizes the holonomy (see
{\cite{Bergou-Wardetzky-EtAl-Discrete-Elastic-Rods-2008}}),
\begin{equation}
  \tmmathbf{k}= \frac{2 \tmmult \tmmathbf{a} \times \tmmathbf{b}}{1
  +\tmmathbf{a} \cdot \tmmathbf{b}} . \label{eq:a-b-to-k}
\end{equation}
The infinitesimal rotation $r_{\tmmathbf{a}} (\delta \theta)$ from
equation~(\ref{eq:variation-parallel-trsp-temp1}) can be found from
equation~(\ref{eq:unit-quaternion-normal-form}) as
\begin{equation}
  \begin{array}{lll}
    r_{\tmmathbf{a}} (\delta \theta) & = & 1 +\tmmathbf{a} \tmmult
    \frac{\delta \theta}{2} +\mathcal{O} (\delta \theta^2)\\
    & = & 1 - \frac{\tmmathbf{k} \cdot \delta \tmmathbf{b}}{4} \tmmult
    \tmmathbf{a}+\mathcal{O} (\delta \theta^2)\\
    & = & 1 - \frac{\tmmathbf{a} \otimes \tmmathbf{k}}{4} \cdot \delta
    \tmmathbf{b}+\mathcal{O} (\delta \theta^2) .
  \end{array} \label{eq:infinitesimal-rotation}
\end{equation}
Equation~(\ref{eq:variation-parallel-trsp-temp1}) is then rewritten with the
help of the operator $\tmmathbf{b}_{\times}$ from
equation~(\ref{eq:cross-operator}) as
\[ \begin{array}{lll}
     p_{\tmmathbf{a}}^{\tmmathbf{b}+ \delta \tmmathbf{b}} & = & \left( 1 +
     \frac{\tmmathbf{b}_{\times}}{2} \cdot \delta \tmmathbf{b} \right) \tmmult
     p_{\tmmathbf{a}}^{\tmmathbf{b}} \tmmult \left( 1 - \frac{\tmmathbf{a}
     \otimes \tmmathbf{k}}{4} \cdot \delta \tmmathbf{b} \right) +\mathcal{O}
     (| \delta \tmmathbf{b} |^2)\\
     & = & \left( 1 + \frac{\tmmathbf{b}_{\times}}{2} \cdot \delta
     \tmmathbf{b}- \frac{(p_{\tmmathbf{a}}^{\tmmathbf{b}} \ast \tmmathbf{a})
     \otimes \tmmathbf{k}}{4} \cdot \delta \tmmathbf{b} \right) \tmmult
     p_{\tmmathbf{a}}^{\tmmathbf{b}} +\mathcal{O} (| \delta \tmmathbf{b}
     |^2)\\
     & = & \left( 1 + \frac{2 \tmmult \tmmathbf{b}_{\times} -\tmmathbf{b}
     \otimes \tmmathbf{k}}{4} \cdot \delta \tmmathbf{b} \right) \tmmult
     p_{\tmmathbf{a}}^{\tmmathbf{b}} +\mathcal{O} (| \delta \tmmathbf{b} |^2)
     .
   \end{array} \]
In view of this, the first order variation of parallel transport writes as
\[ \delta p_{\tmmathbf{a}}^{\tmmathbf{b}} = \frac{1}{2} \tmmult \left( \left(
   \tmmathbf{b}_{\times} - \frac{\tmmathbf{b} \otimes \tmmathbf{k}}{2} \right)
   \cdot \delta \tmmathbf{b} \right) \tmmult p_{\tmmathbf{a}}^{\tmmathbf{b}} .
\]
Identifying with equation~(\ref{eq:quaternion-increment-from-vector}), we find
that it is captured by the infinitesimal rotation vector
\begin{equation}
  \delta \hat{\tmmathbf{p}}_{\tmmathbf{a}}^{\tmmathbf{b}} = \left(
  \tmmathbf{b}_{\times} - \frac{\tmmathbf{b} \otimes \tmmathbf{k}}{2} \right)
  \cdot \delta \tmmathbf{b}.
  \label{eq:generic-variation-of-parallel-trp-first-order}
\end{equation}

\subsubsection{Second variation of parallel transport}

From equation~(\ref{eq:generic-variation-of-parallel-trp-first-order}), we
have
\begin{equation}
  \begin{array}{lll}
    \delta_2 (\delta_1 \hat{\tmmathbf{p}}_{\tmmathbf{a}}^{\tmmathbf{b}}) & = &
    \left( (\delta_2 \tmmathbf{b})_{\times} - \frac{\delta_2 \tmmathbf{b}
    \otimes \tmmathbf{k}+\tmmathbf{b} \otimes \delta_2 \tmmathbf{k}}{2}
    \right) \cdot \delta_1 \tmmathbf{b}+ \left( \tmmathbf{b}_{\times} -
    \frac{\tmmathbf{b} \otimes \tmmathbf{k}}{2} \right) \cdot \delta_{12}
    \tmmathbf{b}\\
    & = & \delta_2 \tmmathbf{b} \times \delta_1 \tmmathbf{b}- \frac{1}{2}
    \delta_2 \tmmathbf{b} \tmmult (\tmmathbf{k} \cdot \delta_1 \tmmathbf{b}) -
    \frac{\tmmathbf{b}}{2} \tmmult \delta_2 \tmmathbf{k} \cdot \delta_1
    \tmmathbf{b}+ \left( \tmmathbf{b}_{\times} - \frac{\tmmathbf{b} \otimes
    \tmmathbf{k}}{2} \right) \cdot \delta_{12} \tmmathbf{b}
  \end{array} \label{eq:d2-parallel-trsprt-tmp1}
\end{equation}
Using equation~(\ref{eq:a-b-to-k}), the variation of the binormal is found as
\[ \begin{array}{lll}
     \delta_2 \tmmathbf{k} & = & \frac{2 \tmmult \tmmathbf{a} \times \delta_2
     \tmmathbf{b}}{1 +\tmmathbf{a} \cdot \tmmathbf{b}} - \frac{2 \tmmult
     \tmmathbf{a} \times \tmmathbf{b}}{(1 +\tmmathbf{a} \cdot \tmmathbf{b})^2}
     \tmmult \tmmathbf{a} \cdot \delta_2 \tmmathbf{b}\\
     & = & \frac{2}{1 +\tmmathbf{a} \cdot \tmmathbf{b}} \tmmult \left(
     \tmmathbf{a} \times \delta_2 \tmmathbf{b}- \frac{\tmmathbf{k}}{2}
     (\tmmathbf{a} \cdot \delta_2 \tmmathbf{b}) \right)\\
     & = & \frac{2}{1 +\tmmathbf{a} \cdot \tmmathbf{b}} \tmmult \left(
     \tmmathbf{a}_{\times} - \frac{\tmmathbf{k} \otimes \tmmathbf{a}}{2}
     \right) \cdot \delta_2 \tmmathbf{b}
   \end{array} \]
Inserting into equation~(\ref{eq:d2-parallel-trsprt-tmp1}) and reordering the
terms, we find
\[ \delta_2 (\delta_1 \hat{\tmmathbf{p}}_{\tmmathbf{a}}^{\tmmathbf{b}}) =
   \delta_2 \tmmathbf{b} \times \delta_1 \tmmathbf{b}+ \left(
   \tmmathbf{b}_{\times} - \frac{\tmmathbf{b} \otimes \tmmathbf{k}}{2} \right)
   \cdot \delta_{12} \tmmathbf{b}- \frac{\tmmathbf{b}}{(1 +\tmmathbf{a} \cdot
   \tmmathbf{b})} \tmmult \left( \delta_1 \tmmathbf{b} \cdot \left(
   \tmmathbf{a}_{\times} - \frac{\tmmathbf{k} \otimes \tmmathbf{a}}{2} \right)
   \cdot \delta_2 \tmmathbf{b} \right) - \frac{\delta_2 \tmmathbf{b} \otimes
   \delta_1 \tmmathbf{b}}{2} \cdot \tmmathbf{k} \]

In view of equation~(\ref{eq:2nd-order-infinitesimal-rotation-practical}), we
can obtain the second-order infinitesimal rotation vector by symmetrizing this
with respect to the increments $\delta_1 \tmmathbf{b}$ and $\delta_2
\tmmathbf{b}$:
\begin{equation}
  \begin{array}{lll}
    \delta_{12} \hat{\tmmathbf{p}}_{\tmmathbf{a}}^{\tmmathbf{b}} & = &
    \frac{\delta_2 (\delta_1 \hat{\tmmathbf{p}}_{\tmmathbf{a}}^{\tmmathbf{b}})
    + \delta_1 (\delta_2
    \hat{\tmmathbf{p}}_{\tmmathbf{a}}^{\tmmathbf{b}})}{2}\\
    & = & \left( \tmmathbf{b}_{\times} - \frac{\tmmathbf{b} \otimes
    \tmmathbf{k}}{2} \right) \cdot \delta_{12} \tmmathbf{b}+ \left( \delta_1
    \tmmathbf{b} \cdot \frac{\tmmathbf{k} \otimes \tmmathbf{a}+\tmmathbf{a}
    \otimes \tmmathbf{k}}{4 \tmmult (1 +\tmmathbf{a} \cdot \tmmathbf{b})}
    \cdot \delta_2 \tmmathbf{b} \right) \tmmult \tmmathbf{b}- \frac{(\delta_1
    \tmmathbf{b} \otimes \delta_2 \tmmathbf{b}+ \delta_2 \tmmathbf{b} \otimes
    \delta_1 \tmmathbf{b})}{2} \cdot \frac{\tmmathbf{k}}{2} .
  \end{array} \label{eq:delta12-p-appendix}
\end{equation}

\subsubsection{Application to a Discrete elastic rod}

In a Discrete elastic rod, the transport is from the undeformed tangent
$\tmmathbf{a}=\tmmathbf{T}^i$ to the deformed tangent
$\tmmathbf{b}=\tmmathbf{t}^i$, see
equation~(\ref{eq:parallel-transport-in-time}). Equation~(\ref{eq:a-b-to-k})
then yields the definition of the binormal $\tmmathbf{k}^i$ announced in
equation~(\ref{eq:variations-ki-def}), and
equation~(\ref{eq:generic-variation-of-parallel-trp-first-order}) yields the
expression for $\delta \hat{\tmmathbf{p}}^i$ announced in
equation~(\ref{eq:variations-p}). In equation~(\ref{eq:delta12-p-appendix}),
condensing the independent variations as $\delta_1 \tmmathbf{b}= \delta_2
\tmmathbf{b}= \delta \tmmathbf{t}^i$ and identifying $\delta_{12}
\hat{\tmmathbf{p}}_{\tmmathbf{a}}^{\tmmathbf{b}} = \delta^2
\hat{\tmmathbf{p}}^i$ and $\delta_{12} \tmmathbf{b}= \delta^2 \tmmathbf{t}^i$
yields the expression of $\delta^2 \hat{\tmmathbf{p}}^i$ announced in
equation~(\ref{eq:variations-p}).

\subsection{Variation of unit tangents}

With $\tmmathbf{E}^i =\tmmathbf{x}_{i + 1} -\tmmathbf{x}_i$ as the segment
vector, the variation of the unit tangent $\tmmathbf{t}^i =\tmmathbf{E}^i / |
\tmmathbf{E}^i |$ from equation~(\ref{eq:adaptation-constraint-current})
writes
\[ \begin{array}{lll}
     \delta \tmmathbf{t}^i & = & \frac{\delta \tmmathbf{E}^i}{|\tmmathbf{E}^i
     |} -\tmmathbf{E}^i \frac{\delta (|\tmmathbf{E}^i |)}{|\tmmathbf{E}^i
     |^2}\\
     & = & \frac{\delta \tmmathbf{E}^i}{|\tmmathbf{E}^i |} -\tmmathbf{E}^i
     \frac{(\tmmathbf{E}^i \cdot \delta \tmmathbf{E}^i)}{|\tmmathbf{E}^i
     |^3}\\
     & = & \frac{\tmmathbf{I}-\tmmathbf{t}^i \otimes
     \tmmathbf{t}^i}{|\tmmathbf{E}^i |} \cdot \delta_1 \tmmathbf{E}^i
   \end{array} \]
With $\delta \tmmathbf{E}^i = \delta \tmmathbf{x}_{i + 1} - \delta
\tmmathbf{x}_i$, this is the expression of the first variation announced in
equation~(\ref{eq:variations-t}).

Next, the second variation is calculated as
\[ \delta_{12} \tmmathbf{t}^i = \left( - \frac{\delta_2 \tmmathbf{t}^i \otimes
   \tmmathbf{t}^i +\tmmathbf{t}^i \otimes \delta_2
   \tmmathbf{t}^i}{|\tmmathbf{E}^i |} - \frac{(\tmmathbf{I}-\tmmathbf{t}^i
   \otimes \tmmathbf{t}^i)}{|\tmmathbf{E}^i |^2} \tmmult \frac{\tmmathbf{E}^i
   \cdot \delta_2 \tmmathbf{E}^i}{|\tmmathbf{E}^i |} \right) \cdot \delta_1
   \tmmathbf{E}^i . \]
Here, we have used $\delta_{12} \tmmathbf{E}^i =\tmmathbf{0}$ since
$\tmmathbf{E}^i =\tmmathbf{x}_{i + 1} -\tmmathbf{x}_i$ depends linearly on the
degrees of freedom. Inserting the expression of the first variations from
equation~(\ref{eq:variations-t}), the second variation $\delta_{12}
\tmmathbf{t}^i$ can be rewritten as

\[ \begin{array}{ccl}
     \delta_{12} \tmmathbf{t}^i & = & \left( -
     \frac{((\tmmathbf{I}-\tmmathbf{t}^i \otimes \tmmathbf{t}^i) \cdot
     \delta_2 \tmmathbf{E}^i) \otimes \tmmathbf{t}^i +\tmmathbf{t}^i \otimes
     ((\tmmathbf{I}-\tmmathbf{t}^i \otimes \tmmathbf{t}^i) \cdot \delta_2
     \tmmathbf{E}^i)}{|\tmmathbf{E}^i |^2} -
     \frac{(\tmmathbf{I}-\tmmathbf{t}^i \otimes
     \tmmathbf{t}^i)}{|\tmmathbf{E}^i |^2} \tmmathbf{t}^i \cdot \delta_2
     \tmmathbf{E}^i \right) \cdot \delta_1 \tmmathbf{E}^i\\
     & = & - \frac{\tau^i_{I \nocomma K \nocomma J} + \tau^i_{J \nocomma K
     \nocomma I} + \tau^i_{I \nocomma J \nocomma K}}{| \tmmathbf{E}^i |^2}
     \tmmult (\delta_1 E^i)_J \tmmult (\delta_2 E^i)_K \tmmult
     \tmmathbf{e}_I\\
     & = & - \frac{((\tmmathbf{\tau}^i)^{T (132)} + (\tmmathbf{\tau}^i)^{T
     (231)} +\tmmathbf{\tau}^i)_{I \nocomma J \nocomma K}}{| \tmmathbf{E}^i
     |^2} \tmmult (\delta_1 E^i)_J \tmmult (\delta_2 E^i)_K \tmmult
     \tmmathbf{e}_I\\
     & = & - \frac{\tmmathbf{\tau}^i + (\tmmathbf{\tau}^i)^{T (132)} +
     (\tmmathbf{\tau}^i)^{T (231)}}{| \tmmathbf{E}^i |^2} : ((\delta_1
     \tmmathbf{x}_{i + 1} - \delta_1 \tmmathbf{x}_i) \otimes (\delta_2
     \tmmathbf{x}_{i + 1} - \delta_2 \tmmathbf{x}_i)),
   \end{array} \]
where the third-order tensor $\tmmathbf{\tau}^i = (\tmmathbf{I}-\tmmathbf{t}^i
\otimes \tmmathbf{t}^i) \otimes \tmmathbf{t}^i$ and its generalized transpose
are defined below equation~(\ref{eq:variations-t}). The expression of
$\delta^2 \tmmathbf{t}^i$ announced in equation~(\ref{eq:variations-t}) is
obtained by condensing $\delta_1 \tmmathbf{x}_i = \delta_2 \tmmathbf{x}_i =
\delta \tmmathbf{x}_i$ and identifying $\delta^2 \tmmathbf{t}^i = \delta_{12}
\tmmathbf{t}^i$.

\subsection{Variation of directors rotation}

In view of equation~(\ref{eq:quaternion-increment-from-vector}), the
infinitesimal rotation vector $\delta \hat{\tmmathbf{d}}^i$ associated with
the directors rotation $d^i$ is
\[ \delta \hat{\tmmathbf{d}}^i = 2 \tmmult \delta d^i \tmmult \overline{d}^i .
\]
Differentiating the expression of $d^i$ from
equation~(\ref{eq:centerline-twist-frame-reconstruction}), we have $\delta d^i
= \delta \left( p^i \tmmult r_{\tmmathbf{T}^i} (\varphi^i) \tmmult D^i \right)
= \delta p^i \tmmult r_{\tmmathbf{T}^i} (\varphi^i) \tmmult D^i + p^i \tmmult
\delta (r_{\tmmathbf{T}^i} (\varphi^i)) \tmmult D^i$.
Equation~(\ref{eq:unit-quaternion-normal-form}) shows that, with a fixed unit
vector $\tmmathbf{T}^i$, $\delta (r_{\tmmathbf{T}^i} (\varphi^i)) =
\frac{1}{2} \tmmult \left[ \delta \varphi^i \tmmult \tmmathbf{T}^i \right]
\tmmult r_{\tmmathbf{T}^i} (\varphi^i)$---here, the vector in square bracket
is an infinitesimal rotation vector, see
equation~(\ref{eq:quaternion-increment-from-vector}). This yields $\delta d^i
= \delta p^i \tmmult r_{\tmmathbf{T}^i} (\varphi^i) \tmmult D^i + \frac{1}{2}
\tmmult p^i \tmmult \delta \varphi^i \tmmult \tmmathbf{T}^i \tmmult
r_{\tmmathbf{T}^i} (\varphi^i) \tmmult D^i$. Inserting into the equation
above, and using $\overline{d}^i = \overline{D}^i \tmmult r_{\tmmathbf{T}^i}
(- \varphi^i) \tmmult \overline{p}^i$ from
equation~(\ref{eq:centerline-twist-frame-reconstruction}), we find
\[ \begin{array}{lll}
     \delta \hat{\tmmathbf{d}}^i & = & \delta \varphi^i \tmmult p^i \tmmult
     \tmmathbf{T}^i \tmmult r_{\tmmathbf{T}^i} (\varphi^i) \tmmult D^i \tmmult
     \overline{d}^i + 2 \tmmult \delta p^i \tmmult r_{\tmmathbf{T}^i}
     (\varphi^i) \tmmult D^i \tmmult \overline{d}^i\\
     & = & \delta \varphi^i \tmmult p^i \tmmult \tmmathbf{T}^i \tmmult
     \overline{p}^i + 2 \tmmult \delta p^i \tmmult \overline{p}^i\\
     & = & \delta \varphi^i \tmmult p^i \ast \tmmathbf{T}^i + \delta
     \hat{\tmmathbf{p}}^i\\
     & = & \delta \varphi^i \tmmult \tmmathbf{t}^i + \delta
     \hat{\tmmathbf{p}}^i,
   \end{array} \]
as announced in equation~(\ref{eq:variations-di}).

The second-order infinitesimal rotation vector is then obtained from
equation~(\ref{eq:2nd-order-infinitesimal-rotation-practical}) as
\[ \begin{array}{lll}
     \delta_{12} \hat{\tmmathbf{d}}^i & = & \frac{1}{2} \tmmult \left(
     \delta_2 \tmmult \left( \delta_1 \varphi^i \tmmult \tmmathbf{t}^i +
     \delta_1 \hat{\tmmathbf{p}}^i \right) + \delta_1 \tmmult \left( \delta_2
     \varphi^i \tmmult \tmmathbf{t}^i + \delta_2 \hat{\tmmathbf{p}}^i \right)
     \right)\\
     & = & \frac{\delta_1 \varphi^i \tmmult \delta_2 \tmmathbf{t}^i +
     \delta_2 \varphi^i \tmmult \delta_1 \tmmathbf{t}^i}{2} + \delta_{12}
     \hat{\tmmathbf{p}}^i .
   \end{array} \]
Here, we have used $\delta_{12} \varphi^i = 0$ as $\varphi^i$ is a degree of
freedom and the variations $\delta_1 \varphi^i$ and $\delta_2 \varphi^i$ are
independent.

Upon condensation of the two variations, the equation leads to the expression
of $\delta^2 \hat{\tmmathbf{d}}^i$ announced in
equation~(\ref{eq:variations-di}).

\subsection{Rotation gradient}

In view of equation~(\ref{eq:quaternion-increment-from-vector}), the
infinitesimal rotation vector $\delta \hat{\tmmathbf{q}}_i$ associated with
the rotation gradient $q_i = \overline{d^{i - 1}} d^i$ from
equation~(\ref{eq:rotation-gradient}) writes
\[ \begin{array}{lll}
     \delta \hat{\tmmathbf{q}}_i & = & 2 \tmmult \delta q_i \tmmult
     \overline{q_i}\\
     &  & \left( \overline{2 \tmmult \delta d^{i - 1}} \tmmult d^i +
     \overline{d^{i - 1}} \tmmult 2 \tmmult \delta d^i \right) \tmmult
     \overline{q_i}\\
     & = & \overline{d^{i - 1}} \tmmult (- \delta \hat{\tmmathbf{d}}^{i - 1}
     + \delta \hat{\tmmathbf{d}}^i) \tmmult d^{i - 1}
   \end{array} \]
as announced in equation~(\ref{eq:variations-qi}).

The following identity yields the variation of the vector $\overline{s} \ast
\tmmathbf{u}$ obtained by applying the inverse $\overline{s}$ of a rotation
$s$ to a vector $\tmmathbf{u}$,
\[ \begin{array}{lll}
     \delta (\overline{s} \ast \tmmathbf{u}) & = & \delta \left( \overline{s}
     \tmmult \tmmathbf{u} \tmmult s \right)\\
     & = & \overline{\delta s} \tmmult \tmmathbf{u} \tmmult s + \overline{s}
     \tmmult \tmmathbf{u} \tmmult \delta s + \overline{s} \tmmult \delta
     \tmmathbf{u} \tmmult s\\
     & = & \frac{- \overline{s} \tmmult \delta \hat{\tmmathbf{s}} \tmmult
     \tmmathbf{u} \tmmult \overline{s} + \overline{s} \tmmult \tmmathbf{u}
     \tmmult \delta \hat{\tmmathbf{s}} \tmmult s}{2} + \overline{s} \ast
     \delta \tmmathbf{u}\\
     & = & \frac{- (\overline{s} \ast \delta \hat{\tmmathbf{s}}) \tmmult
     (\overline{s} \ast \tmmathbf{u}) + (\overline{s} \ast \tmmathbf{u})
     \tmmult (\overline{s} \ast \delta \hat{\tmmathbf{s}})}{2} + \overline{s}
     \ast \delta \tmmathbf{u}\\
     & = & - (\overline{s} \ast \delta \hat{\tmmathbf{s}}) \times
     (\overline{s} \ast \tmmathbf{u}) + \overline{s} \ast \delta \tmmathbf{u}.
   \end{array} \]
With $\delta = \delta_1$, $s = d^{i - 1}$ and $\tmmathbf{u}= \delta_2
\hat{\tmmathbf{d}}^i - \delta_2 \hat{\tmmathbf{d}}^{i - 1}$, we have
$\overline{s} \ast \tmmathbf{u}= \overline{d^{i - 1}} \ast (\delta_2
\hat{\tmmathbf{d}}^i - \delta_2 \hat{\tmmathbf{d}}^{i - 1}) = \delta_2
\hat{\tmmathbf{q}}_i$, see equation~(\ref{eq:variations-qi}), and the identity
above yields
\[ \begin{array}{lll}
     \delta_1 (\delta_2 \hat{\tmmathbf{q}}_i) & = & - (\overline{d^{i - 1}}
     \ast \delta_1 \hat{\tmmathbf{d}}^{i - 1}) \times \delta_2
     \hat{\tmmathbf{q}}_i + \overline{d^{i - 1}} \ast (\delta_1 (\delta_2
     \hat{\tmmathbf{d}}^i) - \delta_1 (\delta_2 \hat{\tmmathbf{d}}^{i - 1}))\\
     & = & \overline{d^{i - 1}} \ast (\delta_1 (\delta_2
     \hat{\tmmathbf{d}}^i) - \delta_1 (\delta_2 \hat{\tmmathbf{d}}^{i - 1})) +
     \delta_2 \hat{\tmmathbf{q}}_i \times (\overline{d^{i - 1}} \ast \delta_1
     \hat{\tmmathbf{d}}^{i - 1})
   \end{array} . \]
Symmetrizing with respect to the independent variations $\delta_1$ and
$\delta_2$ and using
equation~(\ref{eq:2nd-order-infinitesimal-rotation-practical}), we obtain the
second infinitesimal vector as
\[ \delta_{12} \hat{\tmmathbf{q}}_i = \overline{d^{i - 1}} \ast \left(
   \delta_{12} \hat{\tmmathbf{d}}^i {- \delta_{12}}  \hat{\tmmathbf{d}}^{i -
   1} \right) + \frac{\delta_1 \hat{\tmmathbf{q}}_i \times (\overline{d^{i -
   1}} \ast \delta_2 \hat{\tmmathbf{d}}^{i - 1}) + \delta_2
   \hat{\tmmathbf{q}}_i \times (\overline{d^{i - 1}} \ast \delta_1
   \hat{\tmmathbf{d}}^{i - 1})}{2} . \]
Upon condensation of the two variations, the equation leads to the expression
of $\delta^2 \hat{\tmmathbf{q}}_i$ announced in
equation~(\ref{eq:variations-qi}).

\subsection{Strain vector}

Equation~(\ref{eq:kappi}) can be rewritten as $\tmmathbf{\kappa}_i = 2 \tmmult
\mathcal{I} (q_i)$, where $\mathcal{I} (q) = \frac{q - \overline{q}}{2}$
denotes the vector part of a quaternion. The operator $\mathcal{I}$ being
linear, we have
\[ \begin{array}{lll}
     \delta \tmmathbf{\kappa}_i & = & 2 \tmmult \mathcal{I} (\delta q_i)\\
     & = & \mathcal{I} \left( \delta \hat{\tmmathbf{q}}_i \tmmult q_i \right)
   \end{array} \]
as well as
\[ \begin{array}{lll}
     \delta_{12} \tmmathbf{\kappa}_i & = & 2 \tmmult \mathcal{I} (\delta_{12}
     q_i)\\
     & = & \mathcal{I} \left( \Bigl( \delta_{12} \hat{\tmmathbf{q}}_i -
     \frac{\delta_1 \hat{\tmmathbf{q}}_i \cdot \delta_2
     \hat{\tmmathbf{q}}_i}{2} \Bigr) \tmmult q_i \right),
   \end{array} \]
as announced in equation~(\ref{eq:variations-kappai}). In the equation above,
the second variation of the unit quaternion $\delta_{12} q_i$ has been
expressed using equation~(\ref{eq:quaternion-increment-from-vector}).

\subsection{Numerical verification}
We verify the gradient and Hessian of the elastic energy, by
considering a Kirchhoff rod having 80 nodes.  Starting from a straight rod, we increment the magnitude of the natural
curvature, magnitude of gravity, and a point load applied at the ends over 100 iterations. At each iteration we compute the equilibrium,
disabling the update of the reference configuration discussed in
Section~\ref{ssec:reference-vs-current}.  This allows us to verify the
gradient in the generic setting where the reference and current
configurations differ significantly from each other.  The computed
equilibrium solution is denoted by the vector $\mathbf{X}$.  We
introduce a second configuration vector $\tilde{\mathbf{X}}$ by adding
a random perturbation to $\mathbf{X}$ where each perturbation is chosen randomly between $(-0.1,0.1)$. This magnitude of perturbation ensures that the configuration $\tilde{\mathbf{X}}$ is sufficienlty far from an equilibrium. By starting with the different equilibrium solutions $\mathbf{X}$, we ensure that the variations are taken at different locations in the configuration space.

The gradient of the discrete strain energy $\mathcal{E} = \sum_{i=1}^{N-1}E_{i}$ is evaluated at the point $\tilde{\mathbf{X}}$ either as $\nabla \mathcal{E}_{a}$ computed based on the analytical formula given in the main text, or as $\nabla \mathcal{E}_{fd}$ using finite differences as $(\nabla \mathcal{E}_{fd})_i = (\mathcal{E}(\tilde{\mathbf{X}} + h \mathbf{e}_i) - \mathcal{E}(\tilde{\mathbf{X}} + h \mathbf{e}_i))/2h$ where $h = 10^{-7}$ and $\mathbf{e}_i$ is a unit vector where the $i^\text{th}$ component is $1$.

We then calculate the relative gradient error as:
\[
\| \nabla \mathcal{E}_{\textrm{err}} \| = \frac{\| \nabla \mathcal{E}_a - \nabla \mathcal{E}_{fd} \|_\infty}{ \|\nabla  \mathcal{E}_a \|_{\infty}}
\]

Similarly for the Hessian, we calculate the hessian $\nabla^2
\mathcal{E}$ of the strain energy gradient at the point $\tilde{\mathbf{X}}$, either analytically
($\nabla^2 \mathcal{E}_{a}$) using the methods described in the manuscript or using finite differences
($\nabla^2 \mathcal{E}_{fd}$). We calculate $\nabla^2 \mathcal{E}_{fd}$ using finite differences on the analytical form of the gradient. The relative hessian error is calculated as:
\[
\| \nabla^{2}\mathcal{E}_{\textrm{err}} \| = \frac{\| \nabla^2 \mathcal{E}_a - \nabla^2 \mathcal{E}_{fd} \|_\infty}{ \| \nabla^2 \mathcal{E}_a \|_\infty } \, .
\]

\begin{figure}[h]
  \centerline{\includegraphics{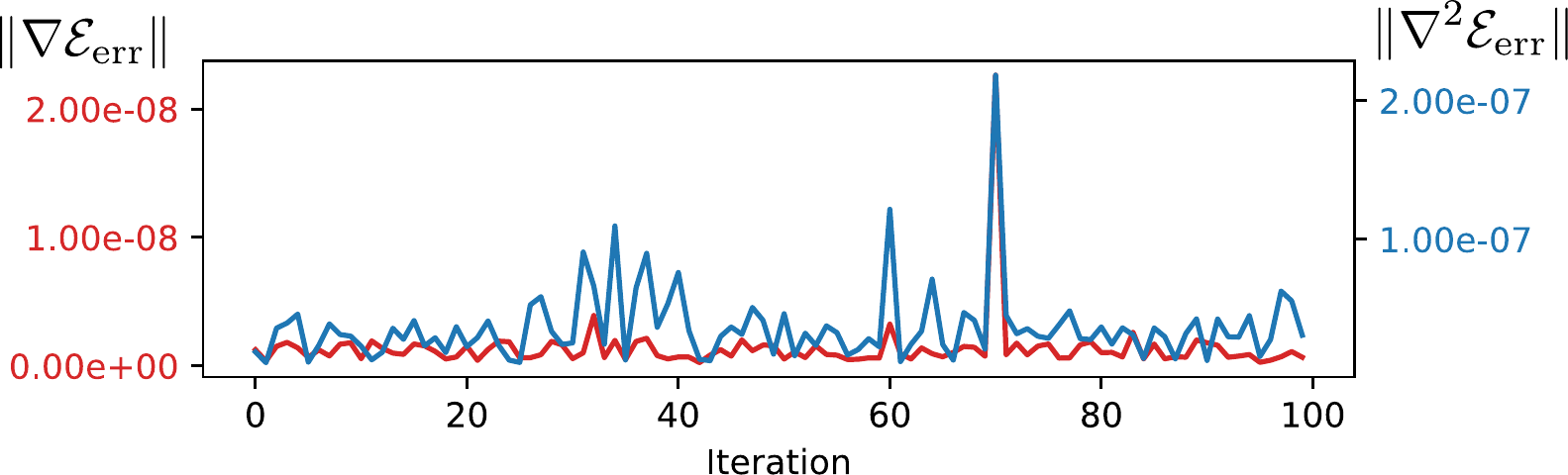}}
  \caption{Errors between the analytical and numerical calculations
  for gradients and hessians.\label{fig:Error}}
\end{figure}

At every iteration, we calculate a different random perturbation and calculate the errors $\|\nabla \mathcal{E}_\textrm{err}\|$ and $\| \nabla^2 \mathcal{E}_{\textrm{err}}\|$ at that point. The results are shown in Figure~(\ref{fig:Error}).

\bibliographystyle{plain}
% \bibliography{mybib}

\end{document}